\documentclass[journal]{IEEEtran}

\usepackage{xcolor,soul,framed} 

\colorlet{shadecolor}{yellow}
\usepackage[pdftex]{graphicx}
\graphicspath{{../pdf/}{../jpeg/}}
\DeclareGraphicsExtensions{.pdf,.jpeg,.png}

\usepackage[cmex10]{amsmath}
\usepackage{array}
\usepackage{mdwmath}
\usepackage{mdwtab}
\usepackage{eqparbox}
\usepackage{url}

\usepackage{amssymb}  

\usepackage{relsize} 

\usepackage[ruled,linesnumbered]{algorithm2e} 
\SetKwRepeat{Do}{do}{while} 
\usepackage{setspace} 
\usepackage{multirow} 
\usepackage{hyperref}
\usepackage{pdfpages}  
\usepackage{subfigure} 
\usepackage{tabularx}  

\def\BibTeX{{\rm B\kern-.05em{\sc i\kern-.025em b}\kern-.08em
    T\kern-.1667em\lower.7ex\hbox{E}\kern-.125emX}}
\begin{document}
\title{EEG-DG: A Multi-Source Domain Generalization Framework for Motor Imagery EEG Classification}

\author{Xiao-Cong Zhong, Qisong Wang, Dan Liu, Zhihuang Chen, Jing-Xiao Liao, Jinwei Sun,~\IEEEmembership{Member,~IEEE,} \\ Yudong Zhang,~\IEEEmembership{Senior Member,~IEEE,} and Feng-Lei Fan,~\IEEEmembership{Member,~IEEE}
\thanks{Manuscript received October XX, XXXX. This work was sponsored by the Fundamental Research Funds for the Central Universities (Grant No. IR2021222), Future Science and Technology Innovation Team project of HIT (216506). \textit{(Corresponding authors: Qisong Wang; Dan Liu; Feng-Lei Fan.)}}
\thanks{Xiao-Cong Zhong, Qisong Wang, Dan Liu, Zhihuang Chen, and Jinwei Sun are with the School of Instrumentation Science and Engineering, Harbin Institute of Technology, Harbin 150001, China (e-mail: zhongxiaocong@hit.edu.cn; wangqisong@hit.edu.cn; liudan@hit.edu.cn; 1190302305@stu.hit.edu.cn; jingxiaoliao@hit.edu.cn; jwsun@hit.edu.cn).}%
\thanks{Yudong Zhang is with School of Computing and Mathematical Sciences, University of Leicester, Leicester, United Kingdom (email: yudong.zhang@le.ac.uk)}
\thanks{Feng-Lei Fan is with Department
of Mathematics, The Chinese University of Hong Kong, Shatin, Hong Kong. (e-mail: flfan@math.cuhk.edu.hk).}
    }

\markboth{Journal of \LaTeX}%
{Zhong \MakeLowercase{\textit{et al.}}: EEG-DG: A Multi-Source Domain Generalization Framework for Motor Imagery EEG Classification}
\maketitle

\begin{abstract}
Motor imagery EEG classification plays a crucial role in non-invasive Brain-Computer Interface (BCI) research. However, the classification is affected by the non-stationarity and individual variations of EEG signals. Simply pooling EEG data with different statistical distributions to train a classification model can severely degrade the generalization performance. To address this issue, the existing methods primarily focus on domain adaptation, which requires access to the target data during training. This is unrealistic in many EEG application scenarios. In this paper, we propose a novel multi-source domain generalization framework called EEG-DG, which leverages multiple source domains with different statistical distributions to build generalizable models on unseen target EEG data.
We optimize both the marginal and conditional distributions to ensure the stability of the joint distribution across source domains and extend it to a multi-source domain generalization framework to achieve domain-invariant feature representation, thereby alleviating calibration efforts. 
Systematic experiments on a simulative dataset and BCI competition datasets IV-2a and IV-2b demonstrate the superiority of our proposed EEG-DG over state-of-the-art methods. 
Specifically, EEG-DG achieves an average classification accuracy/kappa value of 81.79\%/0.7572 and 87.12\%/0.7424 on datasets IV-2a and IV-2b, respectively, which even outperforms some domain adaptation methods. 
Our code is available at \url{https://github.com/XC-ZhongHIT/EEG-DG} for free download and evaluation.
\end{abstract}

\begin{IEEEkeywords}
Brain-computer interface (BCI), Electroencephalography (EEG), Motor imagery (MI), Domain generalization (DG), and Domain-invariant feature representation.
\end{IEEEkeywords}

\section{Introduction}
\label{sec:introduction}
\IEEEPARstart{B}{rain}-computer interface realizes direct information exchange between the brain and the external environment by recording and decoding EEG signals or other electrophysiological activities regarding brain functionalities \cite{wolpaw2002brain}. Currently, BCI techniques have been applied to various scenarios such as stroke treatment \cite{ang2011large}, anesthesia monitoring \cite{park2020real}, alternative communication typing \cite{kundu2022brain} as well as robot control \cite{rebsamen2007controlling, bogue2018rehabilitation}. In particular, the motor imagery-based BCI offers a convenient and efficient approach to controlling external devices by translating the user’s intended action into control signals \cite{pfurtscheller2001motor}, thereby expediting motor rehabilitation for post-stroke patients \cite{berger2019current}. Specifically, when people imagine different body movements, the changes in the electrical activities of the human cortex present different patterns. Such changes can be captured and deciphered to control rehabilitation robots \cite{wierzgala2018most}. To this end, the classification accuracy of EEG signals is vital in determining the overall performance of a BCI system.

Traditional EEG analysis methods rely on machine learning classification algorithms, coupled with hand-crafted features.
Pfurtscheller \textit{et al.} \cite{pfurtscheller1997eeg} proposed event-related synchronization and event-related desynchronization, which have been widely used as indicators of activation or deactivation states of the brain. In \cite{jin2020internal}, a common spatial pattern (CSP) was utilized to compute the optimal spatial filter, which can maximize or minimize the filter variance ratio between different classes to extract spatial features from EEG signals. Afterward, many CSP-based strategies, such as the composite common spatial pattern (CCSP) \cite{kang2009composite}, the filter bank common spatial pattern (FBCSP) \cite{ang2008filter}, and the regularizing common spatial pattern (RCSP) \cite{lotte2010regularizing}, were developed to improve the effectiveness of EEG signal feature extraction. In recent years, as deep learning rapidly evolves and dominates the field of signal processing, considerable efforts have been devoted to translating deep neural networks into the task of EEG classification.
Models like EEGNet \cite{lawhern2018eegnet}, ConvNet \cite{schirrmeister2017deep}, and C2CM \cite{sakhavi2018learning} moved a major stride by demonstrating the superiority of deep learning in feature extraction and end-to-end classification.

Unfortunately, despite achievements made by these arts, there remains an important challenge in front of us that needs to be addressed. Due to the non-stationarity and individual variations \cite{berkhout1968temporal, zhong2023deep}, EEG signals collected from different timestamps and subjects usually exhibit diverse patterns, \textit{i.e.}, different statistical distributions, even though under the same motor imagery conditions. 
Thus, a long period of data collection, training, and calibration is usually required at the beginning of each usage and each new subject, which severely limits the usability and scalability of BCI \cite{wu2020transfer,zhang2020motor}.
However, traditional machine learning methods cannot address the calibration issue because they typically assume that the training and test data have the same statistical distribution. This significantly hurts the models' generalization performance. To overcome this limitation, a direct solution is to pool training data from multiple recordings; nevertheless, the diverse statistical distributions among different subjects and sessions still often lead to suboptimal outcomes. A more crafted solution is domain adaptation (DA) that adapts EEG data with different statistical distributions and achieves significant gains \cite{hong2021dynamic}, \cite{chen2022single}, \cite{huang2023shallow, she2023improved}. Different from traditional machine learning, DA methods train classification models with strong transferability from the training and test data with different statistical distributions. However, it requires test data to be accessed in the training stage \cite{pan2009survey}, which is not applicable for online applications where test data cannot be collected from users in advance. Especially in healthcare, it's impractical to capture EEG data of body movement imagination for training from each and every stroke patient. Furthermore, considering that EEG data involves sensitive health and mental information, this raises privacy leakage concerns \cite{xia2022privacy}. Hereafter, we formally refer to training and test data as the source and target domains, respectively, and use them interchangeably. 

In recent years, domain generalization (DG) aims to construct models by employing the transferable knowledge from multiple domains without access to the target domain, which is an advanced method in terms of learning domain-invariant features \cite{zhou2022domain, wang2022generalizing}. 
Thus, DG can leverage information from multiple source domains in the training stage and meantime enjoy great generalization capability.
So far, DG has been widely used in many important applications such as computer vision \cite{gan2016learning}, \cite{li2018domain}, \cite{dubey2021adaptive, huang2020self}. In this paper, motivated by the effectiveness of DG, we resort to domain generalization techniques to solve the misalignment problem between source and target domains in motor imagery EEG classification. Currently, DG methods can be roughly categorized into three main groups: i) Data augmentation \cite{volpi2018generalizing}, \cite{volpi2019addressing, zhou2020learning}, which involves increasing the quantity and diversity of the source domain through data randomization and generation, is commonly used to prevent overfitting. One drawback of data augmentation techniques is the potential for information distortion or the damage of specific domain features.
ii) Learning strategy \cite{zhou2021domain}, \cite{li2018learning, kim2021selfreg}, which aims to leverage general learning strategies such as ensemble learning, meta-learning, self-supervised learning, and others to promote the generalization capability of models.
However, these approaches typically suffer from high computational complexity, sensitivity to hyperparameters, and lack of interpretability.
iii) Representation learning \cite{muandet2013domain, ganin2016domain, li2018domain}, which learns domain-invariant representations with feature alignment. This approach contributes to alleviating the distribution shift across domains, enabling the model to acquire domain-invariant representations and thereby improving its generalization capability in unseen target domains.


Although domain-invariant feature learning fits the EEG classification, we find that the existing works cannot be directly applied due to the fact that they only consider minimizing the marginal distribution $P(X)$ while ignoring the conditional distribution $P(Y|X)$. However, the conditional distribution $P(Y|X)$ represents important information that pertains to the brain's special biochemical behaviors. According to \cite{janzing2010causal, li2018deep}, in most situations, $P(Y|X)$ can stay the same as $P(X)$ varies when the causal structure is $X \to Y$, where $X$ is the cause of the effect $Y$. However, the causal structure between $Y$ and $X$ is intricate in motor imagery. On the one hand, the causal structure can be $Y \to X$ in motor imagery experiments \cite{barmpas2023improving}, \textit{i.e.}, the motor imagery EEG features are obtained by patients imagining a body movement. On the other hand, the effects of motor imagery can also be influenced by EEG signals \cite{pfurtscheller1992event}. In both cases, $P(Y|X)$ cannot remain stable when $P(X)$ changes, which means that $P(Y|X)$ should be given sufficient attention. In this light, we propose a novel multi-source domain generalization framework (EEG-DG) that can deal with the scenario where both $P(X)$ and $P(Y|X)$ are different across domains and utilize the information of the joint distribution $P(X, Y)$, thereby learning the transferable knowledge for unseen target domains. This is done via capturing the domain-invariant feature representation by minimizing the discrepancy of both $P(X)$ and $P(Y|X)$ across multiple source domains. 
The main contributions of this article are threefold: 
\begin{itemize}
\item We consider a more practical and challenging scenario: domain-generalized motor imagery EEG classification where the target domain EEG data cannot be accessed during the training process.
\item To address the variations in EEG signals across sessions and subjects, we propose a multi-source domain generalization framework (EEG-DG) that learns domain-invariant features with strong representation by optimizing both $P(X)$ and $P(Y|X)$ towards minimizing the discrepancy across a variety of source domains.
\item Systematic experiments on a simulative dataset and two benchmark EEG motor imagery datasets demonstrate that our proposed EEG-DG can deliver superior performance compared to state-of-the-art methods. Particularly, EEG-DG can achieve competitive performance or outperform the domain adaptation methods that can access the target data during the training.
\end{itemize}

\section{Related Works}
\subsection{EEG Classification with Machine Learning}
Traditional EEG classification relies on feature extraction and machine learning to learn EEG decoding rules with high accuracy. For example, Muller \textit{et al.} \cite{muller1999designing} and Ramoser \textit{et al.} \cite{ramoser2000optimal} devised spatial filters for multi-channel EEG to extract discriminative information from the left and right-hand motor imagery classification tasks, which effectively improves the recognition accuracy of motor imagery.
Blankertz \textit{et al.} \cite{blankertz2001classifying} implemented real-time classification of left-right movements of fingers for untrained individuals, which is appealing due to its short response time and high classification accuracy.
Lal \textit{et al.} \cite{lal2004support} leveraged feature selection algorithms combined with the support vector machines (SVM) for EEG channel selection to reduce the number of channels without compromising the classification accuracy.
The feature engineering plus machine learning methods can classify the EEG signals well, while they do not take into account the variations in EEG statistical distribution across subjects and sessions, which hampers the generalization ability of models.

\begin{figure*}[htb]
\centering
\label{fig_framework}
\includegraphics[scale = 0.55,trim=50 5 50 13,clip]{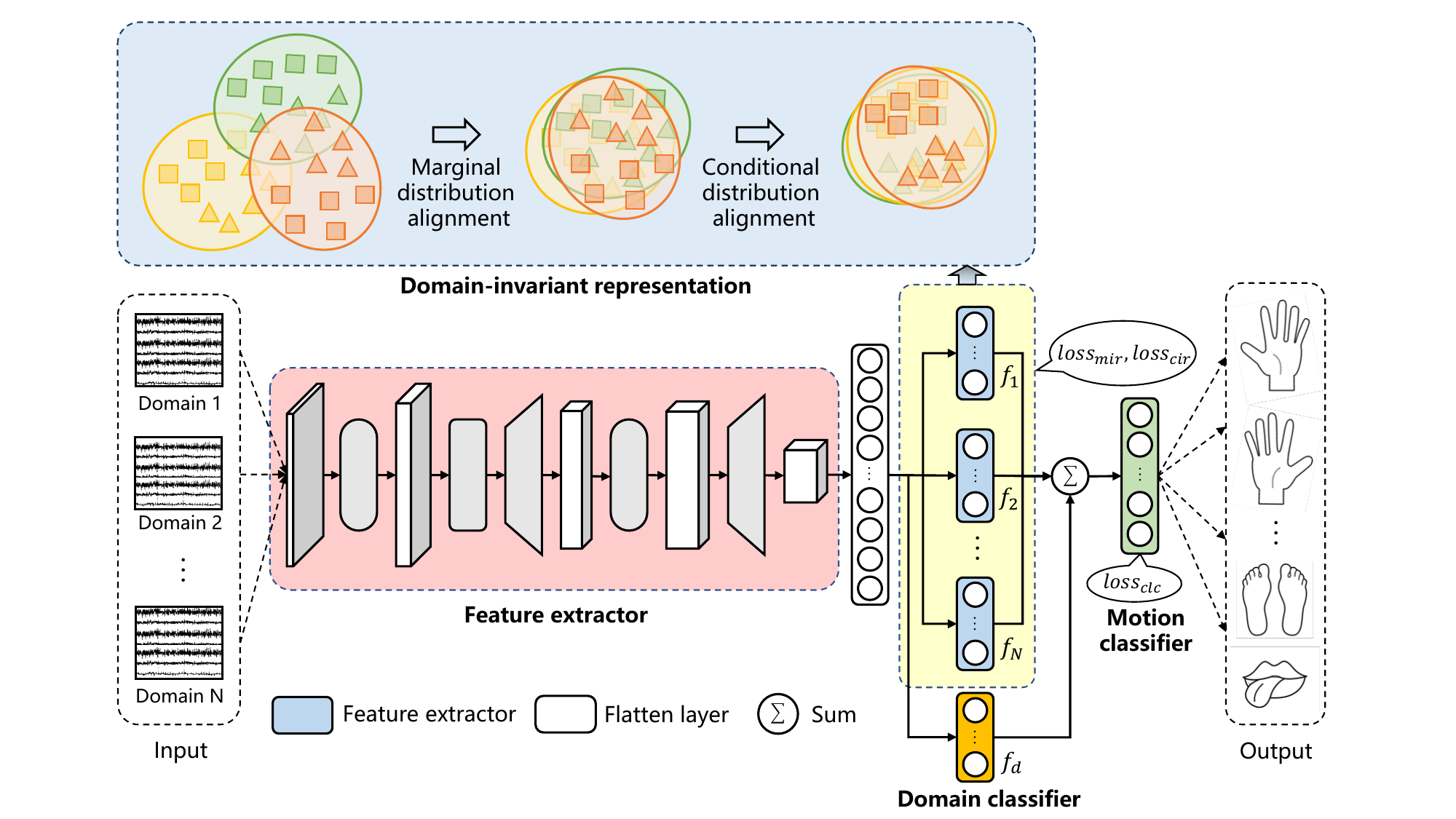}
\caption{The architecture of the proposed EEG-DG framework for motor imagery EEG classification, which consists of four components, including the feature extractor, the domain-invariant representation, the domain classifier, and the motion classifier. The light blue box at the top shows the specific steps of domain-invariant feature representation, where marginal-invariant representation closes the discrepancy of the marginal distributions across source domains, and conditional-invariant representation reduces the difference of the conditional distributions across source domains.} 
\vspace{-0.3cm}
\end{figure*}

\vspace{-0.1cm}
\subsection{Transfer Learning and Domain Adaptation}
Transfer learning describes the procedure of applying models or knowledge in old domains/tasks to new domains/tasks, with domain adaptation being a crucial branch of it (a thorough review of transfer learning, see \cite{pan2009survey}).
In EEG applications, it's well-known that the EEG signal is non-stationary, so in the strict sense, each trial can be regarded as a new task. Recently, many domain adaptation methods tried to find some structure that is invariant across EEG data in both source and target domains \cite{zhao2020deep, phunruangsakao2022deep}, \cite{ zhong2023deep, song2023global}. The covariate shift adaptation was first applied to address the non-stationarity of EEG signals \cite{li2010application}, and it achieved more robust results in the BCI Competition III dataset compared to its competitors.
Hong \textit{et al.} \cite{hong2021dynamic} proposed a dynamic joint domain adaptation network based on adversarial learning to align the marginal distribution across domains and reduce the conditional distribution discrepancy between subdomains.
Chen \textit{et al.} \cite{chen2022single} presented a multi-subdomain adaptive method to solve the shift problem of EEG data and improve classification accuracy.
She \textit{et al.} \cite{she2023improved} introduced an improved domain adaptive network based on the Wasserstein distance that utilizes labeled data from multiple source domains to improve motor imagery classification performance on the target domain. Domain adaptation methods can effectively address the limitation that machine learning methods cannot overcome the variations but require access to the target EEG data during training, which is impractical in many EEG application scenarios.

\vspace{-0.1cm}
\subsection{Domain Generalization}
Domain Generalization learns domain-invariant features from multiple source domains to reduce the prediction error on an unseen target domain.
The primary distinction between DA and DG lies in the fact that DA has access to the target domain during training, whereas DG does not. Therefore, DG is more suitable for EEG classification tasks where collecting data in advance poses significant challenges. 
At present, DG is an emerging field that so far only has a few applications in EEG analysis.
Ma \textit{et al.} \cite{ma2019reducing} a modified deep adversarial network and an adversarial domain generalization framework to address subject variability in EEG-based emotion recognition.
Bethge \textit{et al.} \cite{bethge2022domain} proposed a multi-source learning architecture with private encoders and domain alignment using the maximum mean discrepancy to achieve domain-invariant representations for emotion classification. 

However, few DG works consider optimizing the marginal and conditional distributions of multiple source domains to obtain domain-invariant feature representation, especially in motor imagery EEG classification.
Hence, we strive to exploit domain-invariant feature representation by optimizing the joint distribution of multiple source domains to make predictions on the unseen target domain.

Compared to DA methods, our proposed EEG-DG framework achieves domain-invariant feature representation without relying on any information about the target domain. Instead, it leverages multiple source domains to attain high prediction performance on unseen target domains. This approach not only overcomes the challenge of collecting target domain data in practical applications but also significantly reduces calibration efforts. Additionally, unlike DA methods that utilize pseudo-label strategies for iterative updating to minimize differences in class-conditional distributions $P(X|Y)$, our EEG-DG utilizes the labels of all source domains to ensure the invariance of class-conditional distributions $P(X|Y)$. As a result, it effectively reduces the differences in the joint distribution $P(X, Y)$ across source domains, even in the absence of target data and their labels.

\section{Methodology}
In this section, we first rigorously define the problem of domain-generalized EEG classification. Then we describe the proposed multi-source domain generalization framework for motor imagery EEG classification in detail, with an emphasis on illustrating why the designed loss function is so and how the network is trained. 

\vspace{-0.2cm}
\subsection{Problem Definition and Notations}
Suppose EEG feature and label spaces are represented by $\mathcal{X}$ and $\mathcal{Y}$, respectively. Since data are collected from different scenarios like different subjects or sessions, we suppose to have $N$ well-labeled source domains denoted by $\mathcal{D}_s=\{\mathcal{D}_s^n\}_{n=1}^N$. The $n$-th source domain is denoted as $\mathcal{D}_s^n=\{(\boldsymbol{x}_i^n, y_i^n)\}_{i=1}^{N_s}$, where $\boldsymbol{x}_i^n$ are EEG features, $y_i^n \in \{1,2,\ldots,C\}$ is the label, and $N_s$ is the number of samples. In the experiments, we feed the network with data samples of triple elements: $(\boldsymbol{x}_i, y_i, d_n)$, where $d_n\in \{1, 2, \ldots, N\}$ is augmented to denote that to which domain $(\boldsymbol{x}_i, y_i)$ belongs.
We let an unseen target EEG domain be $\mathcal{D}_t=\{(\boldsymbol{x}_j)\}_{j=1}^{N_t}$. The goal of domain-generalized EEG classification is to learn a domain-generalized classification mapping $f:\mathcal{X} \to \mathcal{Y}$ from source domains $\{\mathcal{D}_s^n\}_{n=1}^N$ that can achieve the minimum prediction error on the unseen target domain $\mathcal{D}_t$.
Mathematically, we train a network by optimizing the following objective function:
\begin{equation}
    f^*= \arg \min \limits_f \frac{1}{N\cdot\!N_s} \sum_{n=1}^{N}\sum_{i=1}^{N_s} \mathcal{L}(f(\boldsymbol{x}_i^n),y_i^n)
\label{eqf}
\end{equation}
where $\mathcal{L}(\cdot,\cdot)$ represents a specific loss function, and $N\cdot\!N_s$ is the number of samples of all source domains, and $f$ is the possible mapping learned by the network. 

\vspace{-0.2cm}
\subsection{Domain-generalized Framework}
Here, we propose a multi-source domain generalization framework for motor imagery EEG classification, called EEG-DG.
An overview of the proposed framework is depicted in Figure \ref{fig_framework}. EEG-DG is composed of four main parts: feature extraction, domain-invariant feature representation, domain classification, and motion classification. The main contribution of our proposed EEG-DG primarily is to guarantee the invariance of both the marginal and conditional distributions across different source domains to learn domain-invariant feature representation that is generalizable.

\subsubsection{Feature Extractor}
\begin{figure}[t]
\centering
\vspace{0.1cm}
\includegraphics[scale = 0.26,trim=0 50 0 45,clip]{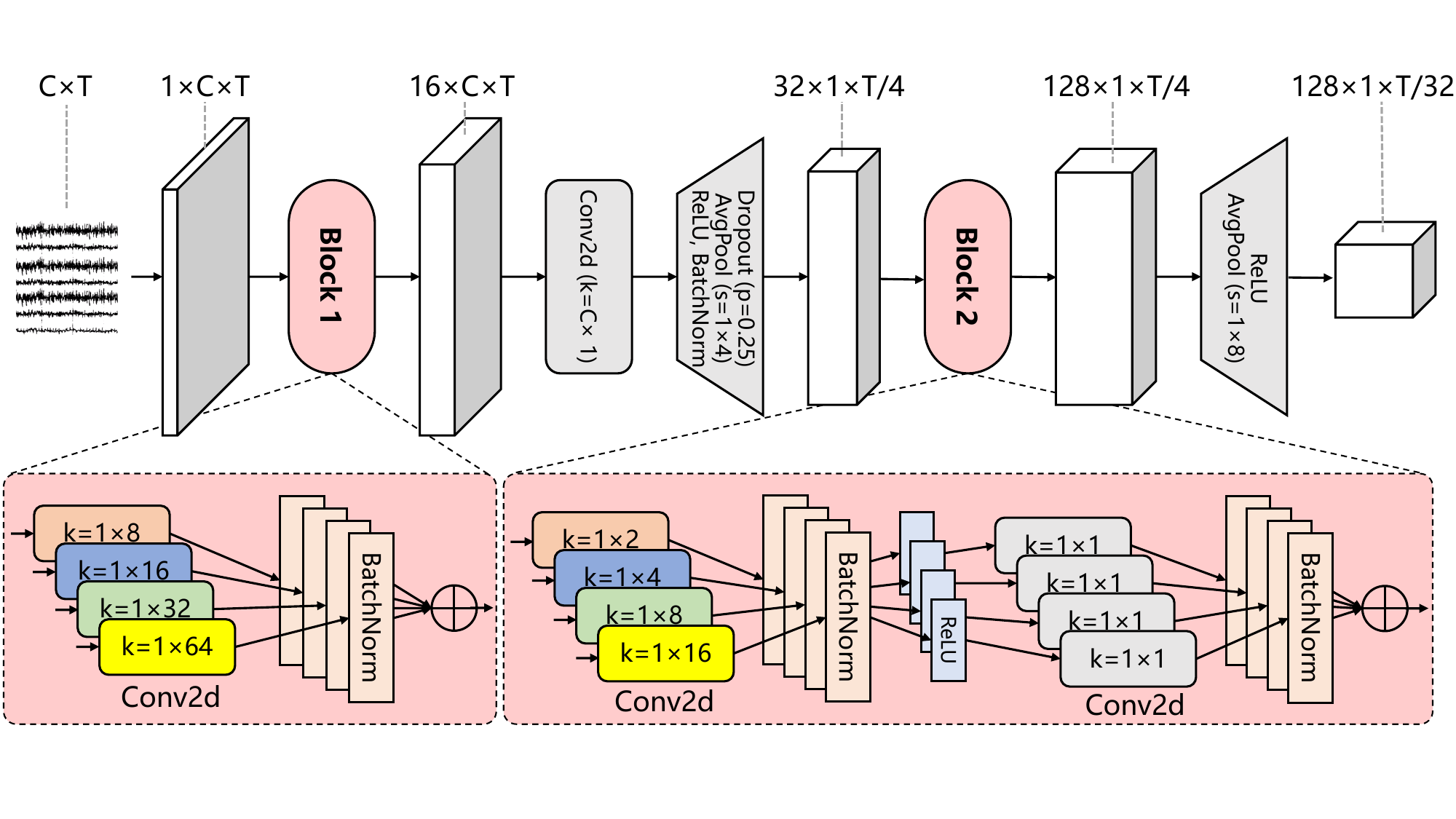}
\caption{Details of the modified EEGNet. The network takes multi-channel EEG data as input $(C, T)$, where $C$ represents the number of channels and $T$ denotes the time dimension.} 
\label{figMS-EEGNet}
\vspace{-0.4cm}
\end{figure}

As depicted in Figure \ref{figMS-EEGNet}, we modify the EEGNet as our backbone. Inspired by the work in \cite{szegedy2017inception}, we integrate the idea of Inception-ResNet to EEGNet to fully exploit deep features by using different-scale kernel functions for motor imagery EEG signals. Specifically, in block 1, four parallel convolutional networks with different kernel sizes are employed to extract temporal features at multiple scales in each EEG electrode. Subsequently, the features are concatenated together. Then, we extract spatial information using a kernel size of $(C\times1)$ and apply batch normalization, exponential linear unit (ELU) activation, dropout, and average pooling, following the same protocol as EEGNet. Additionally, in the separable convolution of block 2, we utilize four parallel convolution operations that are scaled down proportionally to the average-pooled EEG features to obtain more detailed and deeper features. Due to evident differences in temporal features among different subjects and sessions during the motor imagery task, parallel temporal convolutional neural networks are expected to have better capability in generating deep features, compared to the approach with a single kernel.

\subsubsection{Margin-Invariant Feature Representation}
Influenced by the variations of EEG signals, the multi-dimensional EEG features extracted by the above feature extractor still exhibit significant distribution discrepancies. Furthermore, the marginal distribution $P(X)$ reflects general features of EEG signals in specific tasks to some extent \cite{qin2022domain}. To improve the model's generalization performance, here we design a domain-invariant feature representation module that focuses on aligning the marginal distribution across source domains. 

Specifically, we introduce Maximum Mean Discrepancy (MMD) \cite{gretton2006kernel} to measure the similarity of statistical distribution and extend it to multiple domains. Here, the MMD among marginal distributions $ \{P^1(X), P^2(X), . . . , P^N(X)\} $ is defined as the average MMD of each distribution from the average distribution $P_{\bar{X}}$:
\begin{equation}
    \mathrm{avg\_MMD}(P^1_X, P^2_X,..., P^N_X) \!=\! \frac{1}{N}\sum_{n=1}^N \mathrm{MMD}(P^n_X, P_{\bar{X}})
\label{eqavgmmd}
\end{equation}
where $X$ are EEG features, $N$ is the number of source EEG domains, and $P_{\bar{X}}\!=\!\frac{1}{N}\sum_{n=1}^N P_X^n$ represents the average of these statistical distributions of EEG features, \textit{i.e.}, the average geometric center. 
MMD measures the difference by mapping the data to Reproducing Kernel Hilbert Space (RKHS) and then calculating the mean discrepancy therein. Assuming that the mapping function of RKHS is denoted as $\phi(\cdot)$, then MMD can be formalized as
\begin{equation}
\begin{split}
    \! \mathrm{MMD}(P_X^n,P_{\bar{X}})\! 
    =& \| \mathbb{E}_{{\boldsymbol{x}\sim{P_X^n}}}[\phi(\boldsymbol{x}_i^n)] \!- \!\mathbb{E}_{{\boldsymbol{x}\sim P_{\bar{X}}}}[\phi(\boldsymbol{x}_j)] \|^2_{\mathcal{H}} \\
    \approx & \|\frac{1}{N_s}\!\sum_{i=1}^{N_s}\!\phi(\boldsymbol{x}_i^n)\!-\!\frac{1}{N\!\cdot\! N_s}\!\sum_{n=1}^{N}\sum_{j=1}^{N_s}\!\phi(\boldsymbol{x}_j^n)\|^2_\mathcal{H},
\end{split}
\label{eqmmd}
\end{equation}
where the second line of the above equation is an empirical approximation to the first line.
The marginal-invariant representation loss can be obtained by combining \eqref{eqavgmmd} and \eqref{eqmmd}:
\begin{equation}
    \mathcal{L}_{mir}=\!\frac{1}{N}\sum_{n=1}^N \!\|\frac{1}{N_s}\!\sum_{i=1}^{N_s}\!\phi(\boldsymbol{x}_i^n)\!-\!\frac{1}{N\!\cdot\!N_s}\!\sum_{n=1}^{N}\sum_{j=1}^{N_s}\!\phi(\boldsymbol{x}_j^n)\|^2_\mathcal{H}
\label{eq_lmir}
\end{equation}
By optimizing \eqref{eq_lmir}, the marginal distribution discrepancy is close to 0, where it reaches a global minimum at $P^1(\phi(X))= P^2(\phi(X))= \ldots = P^N(\phi(X))=P(\phi(\bar{X}))$, where the learned feature representations on all source domains are completely matched in terms of their marginal distributions.

\subsubsection{Condition-Invariant Feature Representation}
The above margin-invariant feature representation achieves alignment of marginal distributions among multiple source domains but fails to guarantee the invariance of the conditional distribution $P(Y|\phi(X))$ among different source domains \cite{li2018deep}. What's worse, it may even enlarge the difference in conditional distributions during the RKHS mapping process \cite{radovanovic2010hubs}. However, the conditional distribution $P(Y|\phi(X))$ is also important as it contains the intrinsic characteristic due to the brain's biochemical behaviors. Thus, ignoring the conditional distribution will hurt the generalization ability of the model.
To complement the information of conditional distribution, we propose a conditional-invariant representation method. The overall idea is according to Bayes' theorem \cite{db2ff0dd-d9f5-37b6-be2e-7da8e4906218}, $P(Y|X)=P(X|Y)P(Y)/P(X)$, where $P(Y)$ is the prior distribution, which is typically assumed to be constant for simplicity. Therefore, by regularizing that the class-conditional distribution $P(X|Y)$ remains unchanged across domains, the invariance of the conditional distribution $P(Y|X)$ can also be guaranteed. 

Specifically, i) we enforce the model to learn a good embedding, where in each domain samples from the same class are encouraged to be closer while samples from different classes are pushed apart from each other.
Here, we denote $\delta_c^n$ and $\delta_s^n$ as the intra-class compactness and the inter-class separability \cite{hu2015deep} of the $n$-th source domain $\mathcal{D}_s^n=\{(\boldsymbol{x}_i^n, y_i^n)\}_{i=1}^{N_s}$, which are defined as
\begin{equation}
    \delta_c^n=\frac{1}{N_s} \sum_{i=1}^{N_s}\sum_{j=1}^{N_s} \mathrm{dist}(\boldsymbol{x}_i^n,\boldsymbol{x}_j^n)  \cdot \mathbb{I}(y_i^n=y_j^n),
\label{eqdc}
\end{equation}
\begin{equation}
    \delta_s^n=\frac{1}{N_s} \sum_{i=1}^{N_s}\sum_{j=1}^{N_s} \mathrm{dist}(\boldsymbol{x}_i^n,\boldsymbol{x}_j^n)  \cdot \mathbb{I}(y_i^n \neq y_j^n),
\label{eqds}
\end{equation}
where $\mathrm{dist}(\cdot,\cdot)$ is the Euclidean distance, and $\mathbb{I}(\cdot)$ is an indicator function that equals $1$ if the condition is true; otherwise, it equals $0$. ii) We also enforce samples from the same class but different domains to get closer. Denote two different source domains as $\mathcal{D}_s^{n_1}$ and $\mathcal{D}_s^{n_2}$ together. Suppose that $\boldsymbol{h}_c$ is the geometric center of the $c$-th class, which is computed as 
\begin{equation}
\boldsymbol{h}_{c}=\frac{1}{N_{c}}\sum_{i=1}^{N_s} \boldsymbol{x}_i \cdot \mathbb{I}(y_i=c),c\in\{1,2,\ldots,C\},
\end{equation}
where $N_c$ is the number of samples in the $c$-th class, then the intra-center compactness across domains can be calculated as
\begin{equation}
    D^{n_1,n_2}_s = \frac{1}{C} \sum_{c_1=1}^{C} \! \sum_{c_2=1}^{C} \mathrm{dist}(\boldsymbol{h}_{c_1}^{n_1}, \boldsymbol{h}_{c_2}^{n_2}) \cdot \mathbb{I}(c_1 = c_2).
\label{eqDc}
\end{equation}

Reducing the distance between samples of the same class \eqref{eqdc} while increasing the distance between samples of different classes \eqref{eqds} can effectively improve the classification accuracy in each domain.
Nevertheless, it gets too high classification accuracy in a single domain, which tends to cause overfitting in other domains. Hence, we continue the effort to minimize the inter-class distance described in \eqref{eqDc}, so as to facilitate the alignment of the conditional distribution across various domains. 
By combining \eqref{eqdc}, \eqref{eqds}, and \eqref{eqDc}, we formulate the condition-invariant representation loss among multiple source domains as 
\begin{equation}
    \mathcal{L}_{cir}=\sum_{n=1}^{N}\left(\delta_c^n-\alpha \cdot \delta_s^n+\frac{1}{2}\cdot\!\sum_{n'\neq n}D^{n',n}\right),
\label{eql_cir}
\end{equation}
where $\alpha>0$ is a weighting factor to balance the distance.

\vspace{-0.3cm}
\subsection{The Training Process of EEG-DG}
\begin{figure*}[ht]
    \centering
    \label{fig_simulation}
    \subfigure[Original simulation data]{\includegraphics[scale = 0.22,trim=107 70 52 52,clip]{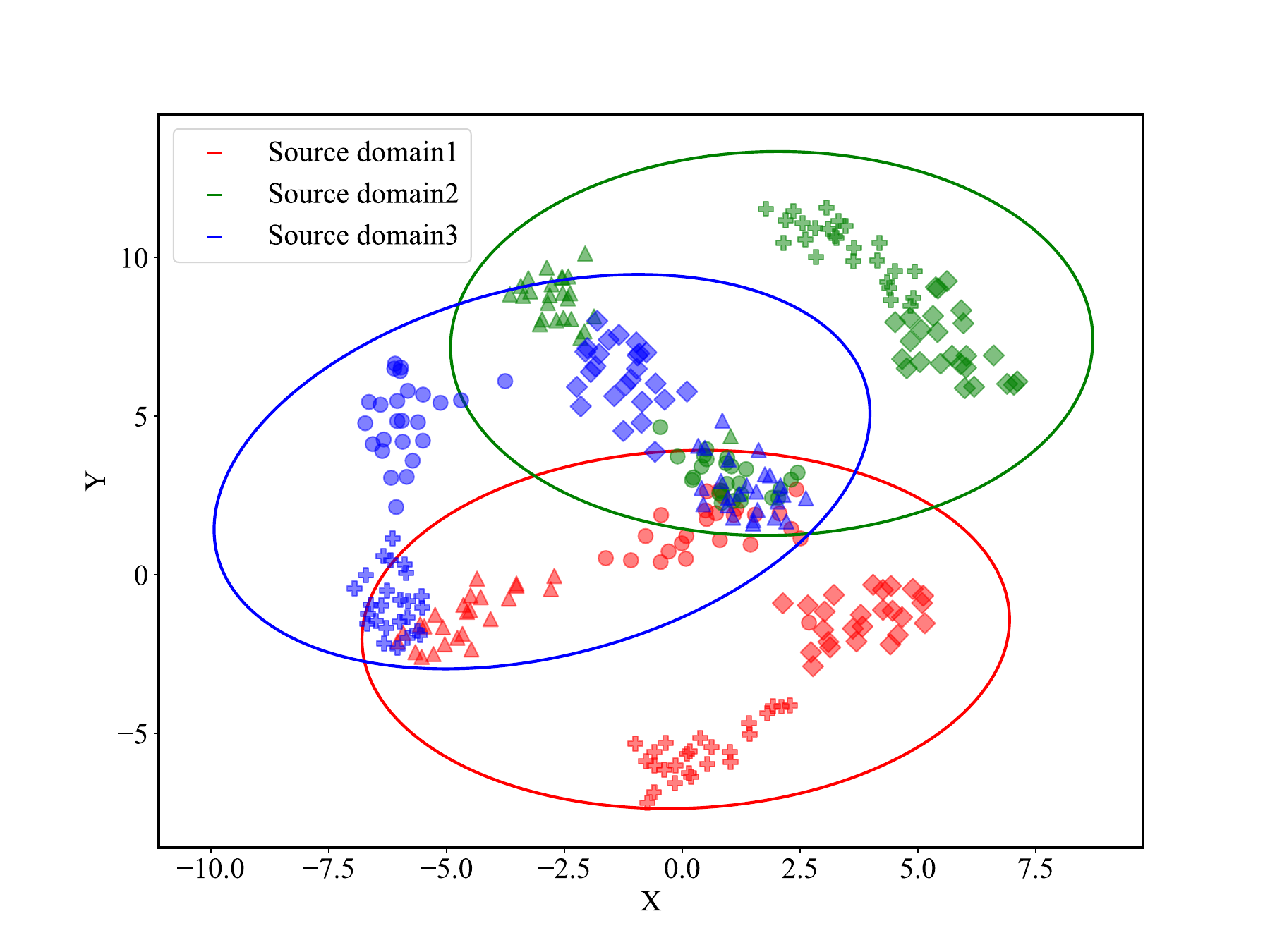}}
    \subfigure[Before EEG-DG]{\includegraphics[scale = 0.22,trim=107 70 52 52,clip]{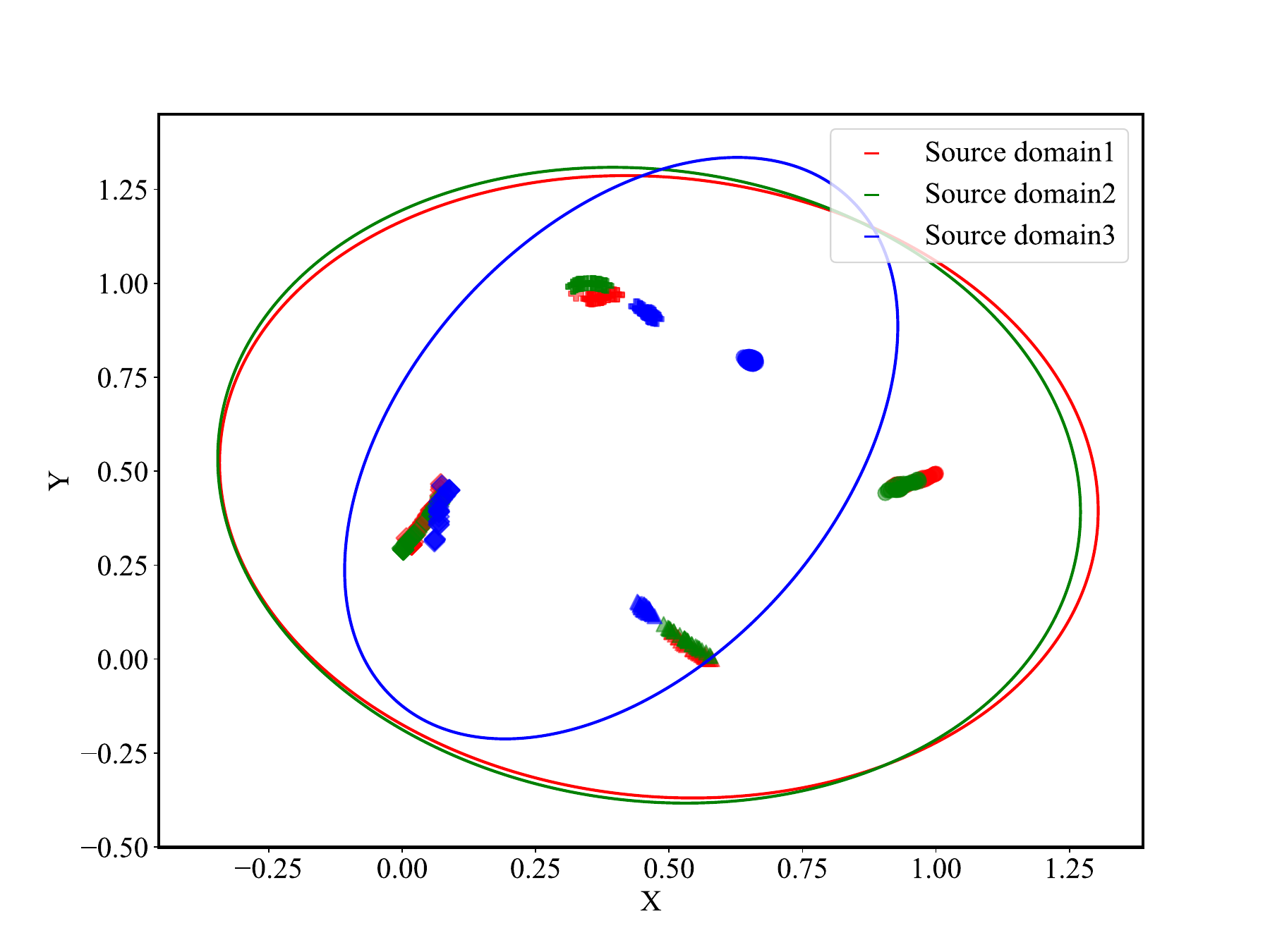}}
    \subfigure[After EEG-DG]{\includegraphics[scale = 0.22,trim=107 70 52 52,clip]{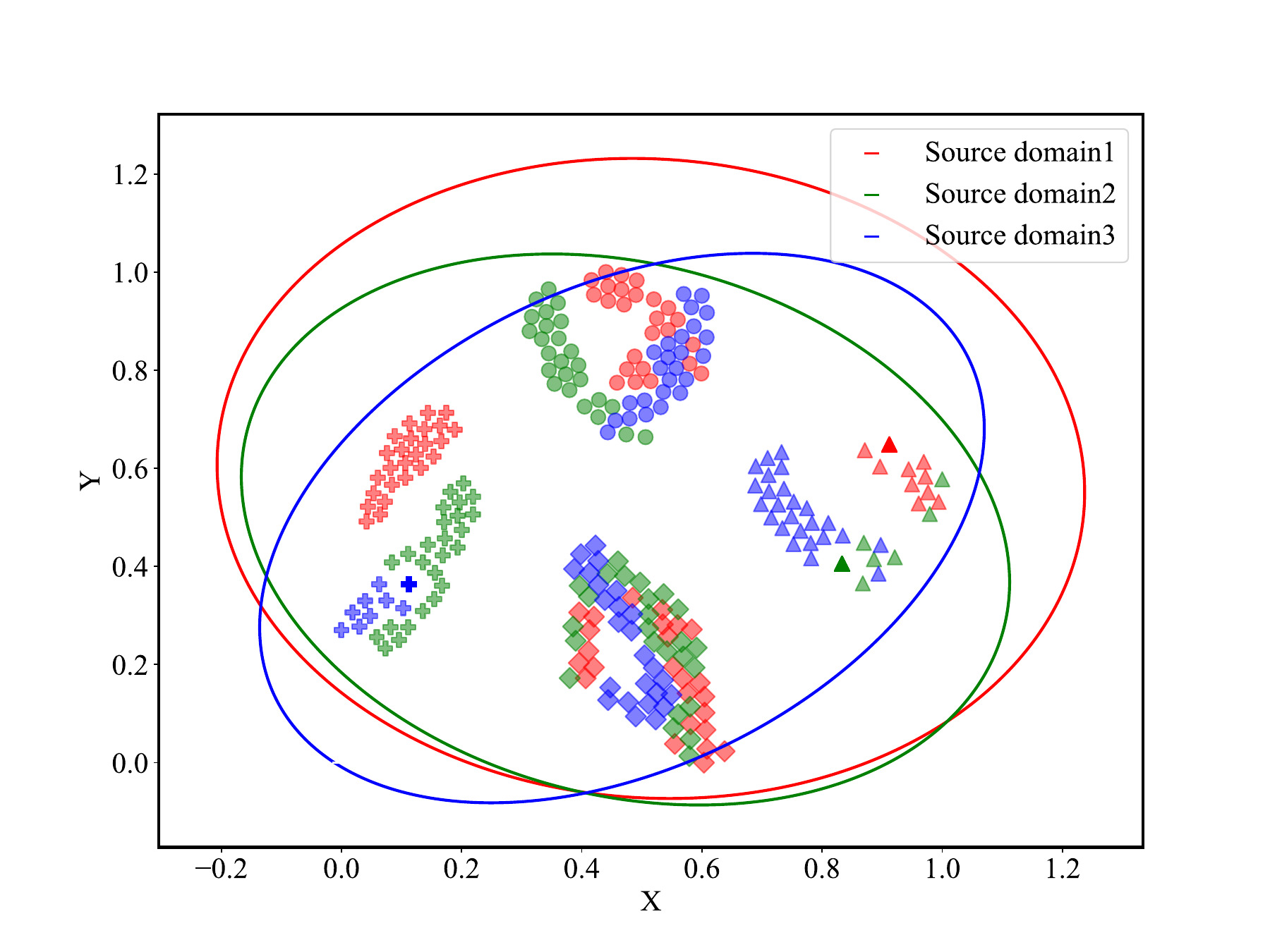}}
    \caption{Statistical distribution visualization with t-SNE of simulation experiments. Different colors represent different source domains, and different shapes represent different classes. The ellipses indicate the confidence ellipses with a confidence probability of 95\%.}
\vspace{-0.3cm}
\end{figure*}

During the training stage, the EEG data of $N$ source domains are first fed simultaneously into the multi-scale feature extractor $g$, where domain-general features can be obtained for each domain. Then, they are passed into each branch, which consists of $K$ fully connected layers $\{f_1,f_2,\ldots,f_N\}$ to extract domain-specific features. Based on these domain-specific features, we seek to learn domain-invariant feature representation by iteratively optimizing the marginal and conditional distribution discrepancy among $N$ source domains, \textit{i.e.}, $\mathcal{L}_{mir}$ and $\mathcal{L}_{cir}$. Additionally, the domain classifier $f_d$ takes domain-general features $g(\boldsymbol{x})$ as the input and produces weights $w_n$ for each domain-specific feature to fuse them in a linear fashion:
\begin{equation}
\boldsymbol{x}_{fused}=\sum_{n=1}^{N} w_n \cdot (f_n \circ g)(\boldsymbol{x}^n),
\label{eqweight}
\end{equation}
where $\sum_{n=1}^{N} w_n = 1$. The fused feature serves as the input to the motion classifier $f_c$ for final classification.
We employ a fully connected layer to perform classification and take the cross entropy loss to train the classifier:
\begin{equation}
    \mathcal{L}_{clc}
    = \frac{1}{N\!\cdot N_s}\! \sum_{n=1}^N \sum\limits_{i=1}^{N_s} J(f_c(\boldsymbol{x}_{fused,i}^n),y_i^n), 
\label{eql_clc}
\end{equation}
where $J(\cdot,\cdot)$ denotes the cross-entropy loss function. Combining the margin-invariant loss, the condition-invariant loss, and the classification loss, the final learning objective of EEG-DG can be formulated as the following optimization problem:
\begin{equation}
    \min \:
    \mathcal{L}=\min \: \mathcal{L}_{clc}+\beta_1 \mathcal{L}_{mir}+\beta_2\mathcal{L}_{cir},
\label{eql}
\end{equation}
where $\beta_1>0$ and $\beta_2>0$ are the tradeoff hyper-parameters. 

\begin{algorithm}[htbp]
\caption{EEG-DG for MI EEG Classification.}
\label{alg:algorithm1}
\setstretch{1.2} 
\KwIn{$N$ source domains \{$\mathcal{D}_s^1$,$\mathcal{D}_s^2$,...,$\mathcal{D}_s^N$\}, and $\beta_1$, $\beta_2$.}
\KwOut{Classification results $\{y_j\}^{N_t}_{j=1}$.} 
Randomly initialize the model parameters $\boldsymbol{p}$\;
\While{not converge}{
    Sample a mini-batch $\mathcal{B}=\{\mathcal{B}_s^1,\ldots,\mathcal{B}_s^N\}_{i=1}^{batch\_size}$\;
    Extract domain-general features $g(\boldsymbol{x})$\;
    Capture domain-specific features $(f_n \circ g)(\boldsymbol{x})$ and produce weights $w_n$ for each source branch\;
    Compute the margin-invariant loss $\mathcal{L}_{mir}$ and the condition-invariant loss $\mathcal{L}_{cir}$ by \eqref{eq_lmir} and \eqref{eql_cir}\;
    Fuse the features $(f_n \circ g)(\boldsymbol{x})$ with $w_n$ by Eq. \eqref{eqweight}\;
    Calculate the classification loss $\mathcal{L}_{clc}$ by \eqref{eql_clc}\;
    Combine $\mathcal{L}_{clc}$, $\mathcal{L}_{mir}$,  and $\mathcal{L}_{cir}$ as $\mathcal{L}$ by \eqref{eql})\;
    Update the parameters $\boldsymbol{p}$ using Adam. }
    Predict labels on the target EEG data $\{\boldsymbol{x}_j\}^{N_t}_{j=1}$. \\
\KwRet Classification results $\{y_j\}^{N_t}_{j=1}$.	
\end{algorithm}

As for the testing, we fix all parameters of the trained network for the target domain. We prepare features of $N+1$ copies that are generated by domain-general feature extractor $g$. Then, we feed $N$ copies into $f_1$,...,$f_N$, respectively, to gain domain-specific features. Subsequently, we feed one copy into the domain classifier to get the weights $w_n$ for each source domain. $w_n$ denotes how similar the target domain is with the $n$-th source domain. Finally, weights and domain-specific features are linearly combined for the final classification. It is worth noting that we do not utilize any information from the target domain in the training process. The complete process of the proposed EEG-DG is summarized in Algorithm \ref{alg:algorithm1}.

\section{Experiments on Simulated Dataset}
To preliminarily validate the performance and present the characteristics of the proposed EEG-DG on multi-source domain alignment, we randomly generated three source domains, each domain consisting of four classes (each class with $25$ samples) but presenting different statistical distributions, as shown in Figure \ref{fig_simulation}(a). 
Different domains are indicated by different colors, and different shapes represent different classes, where the confidence ellipses represent the statistical distributions of the corresponding domains.
First of all, we visualize the effects of data alignment before and after applying the domain-invariant feature representation of EEG-DG on three simulated source domains. The results are depicted in Figure \ref{fig_simulation}(b) and (c), respectively.
Without the domain-invariant feature representation, most samples of the same class become highly clustered but also exist as unclustered samples, such as the blue circular samples. From the view of contrast, samples of the same class from three different source domains are perfectly aligned after the joint distribution alignment. The aligned samples are not simply clustered while maintaining their internal distributional characteristics, which makes the model's decision boundary cleaner.
\begin{table}[t]
\vspace{0.1cm}
\centering
\small
\renewcommand
\arraystretch{1.5}
\caption{Classification Accuracy (in percent $\%$) of Different Methods on Simulated Dataset}
\label{table_simul}
\begin{tabular}{|c|c|c|c|c|c|c|}
\hline
Method & $\mathcal{D}_t^1$ & $\mathcal{D}_t^2$ & $\mathcal{D}_t^3$ & $\mathcal{D}_t^4$ & $\mathcal{D}_t^5$ & MEAN$\pm$STD \\ \hline
SVM & 11.0 & 53.0 & 36.0 & 27.0 & 48.0 & 35.0$\pm$16.8 \\ \hline
$3$NN & 13.0 & 53.0 & 32.0 & 27.0 & 50.0 & 35.0$\pm$16.6 \\ \hline
LDA & 47.0 & 54.0 & 42.0 & 53.0 & 46.0 & 48.4$\pm$5.0 \\ \hline
EEG-DG & \textbf{59.0} & \textbf{88.0} & \textbf{57.0} & \textbf{60.0} & \textbf{64.0} & \textbf{65.6$\pm$12.8} \\ \hline
\end{tabular}
\vspace{-0.4cm}
\end{table}

Furthermore, we randomly generate five target domains with different statistical distributions to evaluate the classification performance. The classification accuracy (MEAN) and standard deviation (STD) of three traditional classifiers, support vector machines (SVM), k-nearest neighbors (kNN), linear discriminant analysis (LDA), and the domain-invariant feature representation are recorded in Table \ref{table_simul}. 
From the results, it is evident that the classification performance of $3$NN and SVM is quite comparable, with LDA showing a slight advantage over both.
In contrast, our proposed EEG-DG method outperforms the traditional classification methods by a significant margin. It achieves an impressive improvement of 30.6\% for 3NN, 30.6\% for SVM, and 17.2\% for LDA, respectively. This remarkable performance further reinforces the notion that our method substantially enhances the model's capability to generalize to unseen target domains. Considering that the target domains we randomly generated exhibit diverse statistical distributions and are more challenging than real-world scenarios, it is acceptable that the overall performance on the four-class classification task is slightly lower.

\section{Experiments on Real-World Dataset}
In this study, we further conduct experiments on two widely used datasets BCI Competition IV Dataset $2$a and $2$b to evaluate the generalization performance of the proposed EEG-DG on the motor imagery classification task and then compare it with other state-of-the-art methods. The experiments confirm that the proposed EEG-DG delivers competitive generalization performance in unseen target domains.

\vspace{-0.2cm}
\subsection{Dataset Discriptions}
\subsubsection{BCI Competition IV Dataset 2a} 
In this dataset, the EEG data of 22 channels were recorded with a sampling rate of 250 Hz from 9 subjects. The participants were instructed to perform four different motor imagery tasks, including the movement of the left hand, right hand, feet, and tongue. Two sessions on different days were collected for each subject, and each session is comprised of 288 trials. In each trial, the participants performed motor imagery from the interval of 2s to 6s. Only one session contains the class labels for all trials, whereas the other was used as the target domain. In this paper, we randomly divide the labeled session into $N$ equal parts as different source domains for training. It is reasonable because there is natural non-stationarity in collected signals across time. 

\subsubsection{BCI Competition IV Dataset 2b}
In Dataset 2b, the EEG data from 3 channels (C3, Cz, and C4) were captured at 250 Hz from 9 subjects. After the cue appeared, all subjects were required to imagine left or right-hand movements for four seconds. Five sessions (each session consisted of 120 trials) were provided for each subject, whereby the first three sessions were well labeled but the last two were not. In the experiments, the first three sessions were hence treated as source domains, and the rest were target domains. In addition, considering that the first three sessions were collected at different times, they naturally serve as 3 different source domains.

\subsection{Experiment Details}
To improve the signal-to-noise ratio, the EEG data was filtered using an 8-35 Hz band-pass filter to obtain frequency bands related to motor imagery. In addition, we standardized the EEG data to [0, 1] with a min-max strategy to make different EEG channels comparable. 
Then, the EEG data was cropped into 4-second windows and fed into the model to reduce the computational effort.
Our proposed EEG-DG was implemented in Python 3.8 using an Intel Core i7 CPU and an NVIDIA RTX 4080 GPU. The framework and loss functions were defined by PyTorch. We trained the model using the Adam optimizer with a learning rate of 0.0005, a mini-batch size of 8, and 500 epochs. We recorded the maximum accuracy for each subject and set parameters $\alpha$, $\beta_1$, and $\beta_2$ to 0.1, 0.1, and 0.1, respectively. The same configuration was applied to both datasets.

To evaluate the proposed EEG-DG, classification accuracy $\mathrm{acc}$ and kappa value $\kappa$ were used to describe the classification performance. 
The kappa value can provide an estimation of the probability of generating accidental results.
\begin{equation}
\label{eq_Acc}
    \mathrm{acc} = \frac{|\boldsymbol{x}_j : \boldsymbol{x}_j \in \mathcal{D}_t \cap \boldsymbol{\hat{y}}_j = \boldsymbol{y}_j|}{| \boldsymbol{x}_j : \boldsymbol{x}_j \in \mathcal{D}_t |} 
\end{equation}
\begin{equation}
\label{eq_kappa}
    \kappa = \frac{\mathrm{acc}-\mathrm{acc_{rand}}}{1-\mathrm{acc_{rand}}}
\end{equation}
where $\boldsymbol{\hat{y}}_j$ is the predicted label, $\boldsymbol{y}_j$ is the true label, and $\mathrm{acc_{rand}}$ denotes the random classification accuracy. 

\subsection{Comparison Methods}
\begin{table*}[htbp]
\centering
\small
\renewcommand\arraystretch{1.5}
\caption{Classification Performance of Different Methods on BCI Competition IV Dataset 2a}
\label{table_2a}
\begin{tabular}{|c|ccccccccc|c|c|c|}
\hline
\multirow{2}{*}{Method} & \multicolumn{9}{c|}{Subject} & \multirow{2}{*}{MEAN$\pm$STD} & \multirow{2}{*}{Kappa} & \multirow{2}{*}{Rank} \\ 
\cline{2-10} & \multicolumn{1}{c|}{A01}  & \multicolumn{1}{c|}{A02}  & \multicolumn{1}{c|}{A03}   & \multicolumn{1}{c|}{A04}  & \multicolumn{1}{c|}{A05}   & \multicolumn{1}{c|}{A06}   & \multicolumn{1}{c|}{A07}   & \multicolumn{1}{c|}{A08}   & A09   &   &  & \\ \hline

FBCSP \cite{ang2008filter}  & \multicolumn{1}{c|}{76.00} & \multicolumn{1}{c|}{56.50} & \multicolumn{1}{c|}{81.25} & \multicolumn{1}{c|}{61.00} & \multicolumn{1}{c|}{55.00} & \multicolumn{1}{c|}{45.25} & \multicolumn{1}{c|}{82.75} & \multicolumn{1}{c|}{81.25} & 70.75 & 67.75$\pm$13.73   & 0.5700 & 9.6 \\ \hline
CCSP \cite{kang2009composite}  & \multicolumn{1}{c|}{84.72} & \multicolumn{1}{c|}{52.78} & \multicolumn{1}{c|}{80.90} & \multicolumn{1}{c|}{59.38} & \multicolumn{1}{c|}{54.51} & \multicolumn{1}{c|}{49.31} & \multicolumn{1}{c|}{88.54} & \multicolumn{1}{c|}{71.88} & 56.60 & 66.51$\pm$15.13                   & 0.5535 & 9.4  \\ \hline
EEGNet \cite{lawhern2018eegnet} & \multicolumn{1}{c|}{79.86} & \multicolumn{1}{c|}{58.68} & \multicolumn{1}{c|}{89.93} & \multicolumn{1}{c|}{64.93} & \multicolumn{1}{c|}{63.19} & \multicolumn{1}{c|}{58.68} & \multicolumn{1}{c|}{64.24} & \multicolumn{1}{c|}{73.61} & 77.08 & 70.22$\pm$10.72                   & 0.6633  & 7.7   \\ \hline
ConvNet \cite{schirrmeister2017deep} & \multicolumn{1}{c|}{76.39} & \multicolumn{1}{c|}{55.21} & \multicolumn{1}{c|}{89.24} & \multicolumn{1}{c|}{74.65} & \multicolumn{1}{c|}{56.94} & \multicolumn{1}{c|}{54.17} & \multicolumn{1}{c|}{92.71} & \multicolumn{1}{c|}{77.08} & 76.39 & 72.53$\pm$14.24  & 0.6338  & 7.4               \\ \hline
C2CM \cite{sakhavi2018learning}  & \multicolumn{1}{c|}{87.50} & \multicolumn{1}{c|}{\textbf{65.28}} & \multicolumn{1}{c|}{90.28} & \multicolumn{1}{c|}{66.67} & \multicolumn{1}{c|}{62.50} & \multicolumn{1}{c|}{45.49} & \multicolumn{1}{c|}{89.58} & \multicolumn{1}{c|}{83.33} & 79.51 & 74.46$\pm$15.33   & 0.6596  & 5.3              \\ \hline \hline
DRDA \cite{zhao2020deep}  & \multicolumn{1}{c|}{83.19} & \multicolumn{1}{c|}{55.14} & \multicolumn{1}{c|}{87.43} & \multicolumn{1}{c|}{75.28} & \multicolumn{1}{c|}{62.29} & \multicolumn{1}{c|}{57.15} & \multicolumn{1}{c|}{86.18} & \multicolumn{1}{c|}{83.61} & 82.00 & 74.75$\pm$12.96                         & 0.6633  & 6.8               \\ \hline
DJDAN \cite{hong2021dynamic} & \multicolumn{1}{c|}{85.77} & \multicolumn{1}{c|}{63.25} & \multicolumn{1}{c|}{93.41} & \multicolumn{1}{c|}{76.75} & \multicolumn{1}{c|}{62.68} & \multicolumn{1}{c|}{\textbf{69.77}} & \multicolumn{1}{c|}{87.37} & \multicolumn{1}{c|}{86.72} & 85.61 & 79.03$\pm$11.35  & 0.7204 & 3.7  \\ \hline
DAFS \cite{phunruangsakao2022deep}  & \multicolumn{1}{c|}{81.94} & \multicolumn{1}{c|}{64.58} & \multicolumn{1}{c|}{88.89} & \multicolumn{1}{c|}{73.61} & \multicolumn{1}{c|}{\textbf{70.49}} & \multicolumn{1}{c|}{56.60} & \multicolumn{1}{c|}{85.42} & \multicolumn{1}{c|}{79.51} & 81.60 & 75.85$\pm$10.47  & 0.6780   & 6.1              \\ \hline
DAWD \cite{she2023improved} & \multicolumn{1}{c|}{83.29} & \multicolumn{1}{c|}{63.97} & \multicolumn{1}{c|}{90.30} & \multicolumn{1}{c|}{76.94} & \multicolumn{1}{c|}{69.34} & \multicolumn{1}{c|}{60.08} & \multicolumn{1}{c|}{89.31} & \multicolumn{1}{c|}{82.35} & 82.81 & 77.60$\pm$10.85                   & 0.6951  & 4.0               \\ \hline
GAT \cite{song2023global} & \multicolumn{1}{l|}{88.89} & \multicolumn{1}{l|}{61.11} & \multicolumn{1}{l|}{93.40} & \multicolumn{1}{l|}{71.86} & \multicolumn{1}{l|}{50.35} & \multicolumn{1}{l|}{60.07} & \multicolumn{1}{l|}{89.58} & \multicolumn{1}{l|}{87.50} & 86.46 & 76.58$\pm$15.98                  & 0.6877   & 4.4              \\ \hline \hline
Our Method & \multicolumn{1}{c|}{\textbf{89.24}}      & \multicolumn{1}{c|}{64.93}      & \multicolumn{1}{c|}{\textbf{94.79}}      & \multicolumn{1}{c|}{\textbf{85.76}}      & \multicolumn{1}{c|}{68.75}      & \multicolumn{1}{c|}{61.46}      & \multicolumn{1}{c|}{\textbf{95.14}}      & \multicolumn{1}{c|}{\textbf{88.89}}      & \textbf{87.15}      &        \textbf{81.79$\pm$13.06}                       &        \textbf{0.7572}       & \textbf{1.4}        \\ \hline
\end{tabular}
\end{table*}

\begin{table*}[htbp]
\centering
\small
\renewcommand\arraystretch{1.5}
\caption{Classification Performance of Different Methods on BCI Competition IV Dataset 2b}
\label{table_2b}
\begin{tabular}{|c|ccccccccc|c|c|c|}
\hline
\multirow{2}{*}{Method} & \multicolumn{9}{c|}{Subject}                                                                                                                                                                                                                  & \multirow{2}{*}{MEAN$\pm$STD} & \multirow{2}{*}{Kappa} & \multirow{2}{*}{Rank} \\ \cline{2-10}
                         & \multicolumn{1}{c|}{B01}   & \multicolumn{1}{c|}{B02}   & \multicolumn{1}{c|}{B03}   & \multicolumn{1}{c|}{B04}   & \multicolumn{1}{c|}{B05}   & \multicolumn{1}{c|}{B06}   & \multicolumn{1}{c|}{B07}   & \multicolumn{1}{c|}{B08}   & B09   &                               &   &                     \\ \hline
FBCSP \cite{ang2008filter}                   & \multicolumn{1}{l|}{70.00} & \multicolumn{1}{l|}{60.36} & \multicolumn{1}{l|}{60.94} & \multicolumn{1}{l|}{97.50} & \multicolumn{1}{l|}{93.12} & \multicolumn{1}{l|}{80.63} & \multicolumn{1}{l|}{78.13} & \multicolumn{1}{l|}{92.50} & 86.88 & 80.00$\pm$13.85                  & 0.6000    &    7.1         \\ \hline
CCSP \cite{kang2009composite}                    & \multicolumn{1}{l|}{63.75} & \multicolumn{1}{l|}{56.79} & \multicolumn{1}{l|}{50.00} & \multicolumn{1}{l|}{93.44} & \multicolumn{1}{l|}{65.63} & \multicolumn{1}{l|}{81.25} & \multicolumn{1}{l|}{72.81} & \multicolumn{1}{l|}{87.81} & 82.81 & 72.70$\pm$14.72                  & 0.4540     &  9.3           \\ \hline
EEGNet \cite{lawhern2018eegnet}                  & \multicolumn{1}{l|}{70.31} & \multicolumn{1}{l|}{70.36} & \multicolumn{1}{l|}{78.44} & \multicolumn{1}{l|}{95.33} & \multicolumn{1}{l|}{93.44} & \multicolumn{1}{l|}{82.18} & \multicolumn{1}{l|}{91.88} & \multicolumn{1}{l|}{87.19} & 71.65 & 82.37$\pm$10.15                   & 0.6507     &     6.0        \\ \hline
ConvNet \cite{schirrmeister2017deep}                 & \multicolumn{1}{l|}{76.56} & \multicolumn{1}{l|}{50.00} & \multicolumn{1}{l|}{51.56} & \multicolumn{1}{l|}{96.88} & \multicolumn{1}{l|}{93.13} & \multicolumn{1}{l|}{85.31} & \multicolumn{1}{l|}{83.75} & \multicolumn{1}{l|}{91.56} & 85.62 & 79.37$\pm$17.26                   & 0.5875     &     6.8        \\ \hline 
C2CM \cite{sakhavi2018learning}  & \multicolumn{1}{c|}{--} & \multicolumn{1}{c|}{--} & \multicolumn{1}{c|}{--} & \multicolumn{1}{c|}{--} & \multicolumn{1}{c|}{--} & \multicolumn{1}{c|}{--} & \multicolumn{1}{c|}{--} & \multicolumn{1}{c|}{--} & \multicolumn{1}{c|}{--} & \multicolumn{1}{c|}{--} & \multicolumn{1}{c|}{--} & \multicolumn{1}{c|}{--} \\

\hline \hline
DRDA \cite{zhao2020deep}                    & \multicolumn{1}{l|}{81.37} & \multicolumn{1}{l|}{62.86} & \multicolumn{1}{l|}{63.63} & \multicolumn{1}{l|}{95.94} & \multicolumn{1}{l|}{93.56} & \multicolumn{1}{l|}{88.19} & \multicolumn{1}{l|}{85.00} & \multicolumn{1}{l|}{\textbf{95.25}} & 90.00 & 83.98$\pm$12.67          & 0.6796      &   4.7         \\ \hline
DJDAN \cite{hong2021dynamic}  & \multicolumn{1}{l|}{75.88} & \multicolumn{1}{l|}{58.57} & \multicolumn{1}{l|}{73.04} & \multicolumn{1}{l|}{96.70} & \multicolumn{1}{l|}{\textbf{98.90}} & \multicolumn{1}{l|}{87.65} & \multicolumn{1}{l|}{85.78} & \multicolumn{1}{l|}{84.35} & 85.31 & 84.66$\pm$12.37                         &  0.6931    &       5.6            \\ \hline
DAFS \cite{phunruangsakao2022deep}                    & \multicolumn{1}{l|}{70.31} & \multicolumn{1}{l|}{\textbf{73.57}} & \multicolumn{1}{l|}{\textbf{80.31}} & \multicolumn{1}{l|}{94.69} & \multicolumn{1}{l|}{95.00} & \multicolumn{1}{l|}{83.75} & \multicolumn{1}{l|}{\textbf{93.73}} & \multicolumn{1}{l|}{95.00} & 75.31 & 84.63$\pm$10.20   & 0.6926 &       4.3          \\ \hline
DAWD \cite{she2023improved}                    & \multicolumn{1}{l|}{\textbf{84.66}} & \multicolumn{1}{l|}{66.57} & \multicolumn{1}{l|}{68.04} & \multicolumn{1}{l|}{96.78} & \multicolumn{1}{l|}{94.32} & \multicolumn{1}{l|}{82.61} & \multicolumn{1}{l|}{88.47} & \multicolumn{1}{l|}{93.96} & 90.10 & 85.06$\pm$11.05                   & 0.7012    &    4.2          \\ \hline
GAT \cite{song2023global}                    & \multicolumn{1}{l|}{84.58} & \multicolumn{1}{l|}{61.67} & \multicolumn{1}{l|}{60.83} & \multicolumn{1}{l|}{\textbf{99.58}} & \multicolumn{1}{l|}{87.50} & \multicolumn{1}{l|}{\textbf{93.33}} & \multicolumn{1}{l|}{85.42} & \multicolumn{1}{l|}{95.00} & \textbf{92.08} & 84.44$\pm$13.98         & 0.6889    &  4.0            \\ \hline \hline
Our Method                    & \multicolumn{1}{l|}{82.50}      & \multicolumn{1}{l|}{67.50}      & \multicolumn{1}{l|}{72.19}      & \multicolumn{1}{l|}{98.44}      & \multicolumn{1}{l|}{96.56}      & \multicolumn{1}{l|}{90.94}      & \multicolumn{1}{l|}{89.38}      & \multicolumn{1}{l|}{95.00}      & 91.56     &            \textbf{87.12$\pm$10.89}                   &  \textbf{0.7424}    &     \textbf{2.6}              \\ \hline 
\end{tabular}
\vspace{-0.2cm}
\end{table*}

To demonstrate the effectiveness of the proposed EEG-DG, we compare the performance of our method with other state-of-the-art methods, including traditional machine learning methods (CCSP \cite{kang2009composite}, FBCSP \cite{ang2008filter}), deep learning models (EEGNet \cite{lawhern2018eegnet}, ConvNet \cite{schirrmeister2017deep}, C2CM \cite{sakhavi2018learning}), and deep domain adaptation methods (DRDA \cite{zhao2020deep},
DJDAN \cite{hong2021dynamic}, DAFS \cite{phunruangsakao2022deep}, DAWD \cite{she2023improved}, and GAT \cite{song2023global}). All these methods are either classic or the latest advances published in this field. 
Note that traditional machine learning methods and deep learning methods do not have access to the target data during training, while domain adaptation methods necessitate access to the target data.
These algorithms are described below: 
\begin{itemize}
\item \textbf{CCSP}:
A modified method of CSP for subject-to-subject transfer, which exploits a linear combination of  covariance matrices of subjects.
\item  \textbf{FBCSP}:
A classification method for EEG signals that extracts CSP features by combining frequency band segmentation and autonomous feature selection.
\item  \textbf{EEGNet}:
A compact convolutional neural network designed for EEG signals. It decodes signals with depthwise and separable convolutions. 
\item \textbf{ConvNet}:
A deep learning model with convolutional neural networks for EEG analysis, which generates the spectral power feature of different frequency bands to decode task-related information.
\item \textbf{C2CM}:
A classification framework developed for the motor imagery EEG data, which utilizes a temporal representation generated by modifying the FBCSP and a convolutional neural network for classification.
\item \textbf{DRDA}:
A domain adaptation method for reducing the intra-subject non-stationarity by optimizing a center loss and matching the feature distribution shift by an adversarial learning strategy.  
\item \textbf{DJDAN}:
A dynamic joint domain adaptation network based on adversarial learning to align the marginal distribution across domains and reduce the conditional distribution discrepancy between sub-domains.
\item \textbf{DAFS}:
An integration method of deep domain adaptation with few-shot learning by leveraging knowledge from multiple subjects to enhance the classification performance of a single subject.
\item \textbf{DAWD}:
An improved domain adaptation network adopting the Wasserstein metric to measure the distance between the source and target domains and aligning the data distributions via an adversarial learning strategy.  
\item \textbf{GAT}:
A novel domain adaptation method based on an attention-based adapter for capturing global correlated features between the source and target domains. A discriminator and an adaptive center loss are designed to reduce the discrepancy between the marginal distribution and the conditional distribution.
\end{itemize}

In this experiment, we either present the best results from published articles if evaluated on datasets IV-$2$a and IV-$2$b or employ their methods on the same datasets with the best hyperparameters reported in those articles. Meanwhile, all methods follow the same data division rules as the competition.  

\begin{figure*}[t]
    \centering
    \subfigure[Subject A03 before EEG-DG]{\includegraphics[scale = 0.22,trim=50 40 50 50,clip]{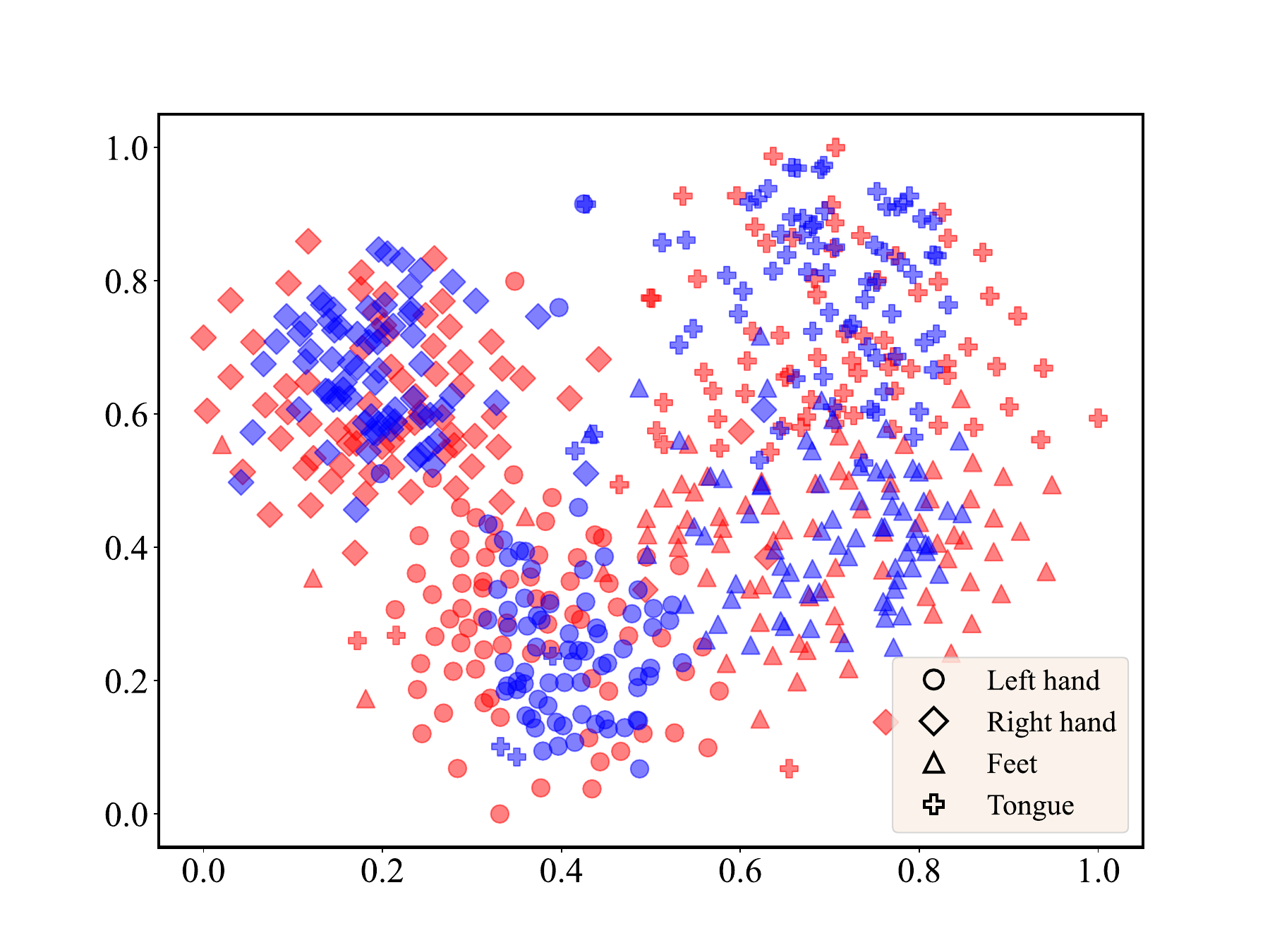}}
    \subfigure[Subject A07 before EEG-DG]{\includegraphics[scale = 0.22,trim=50 40 50 50,clip]{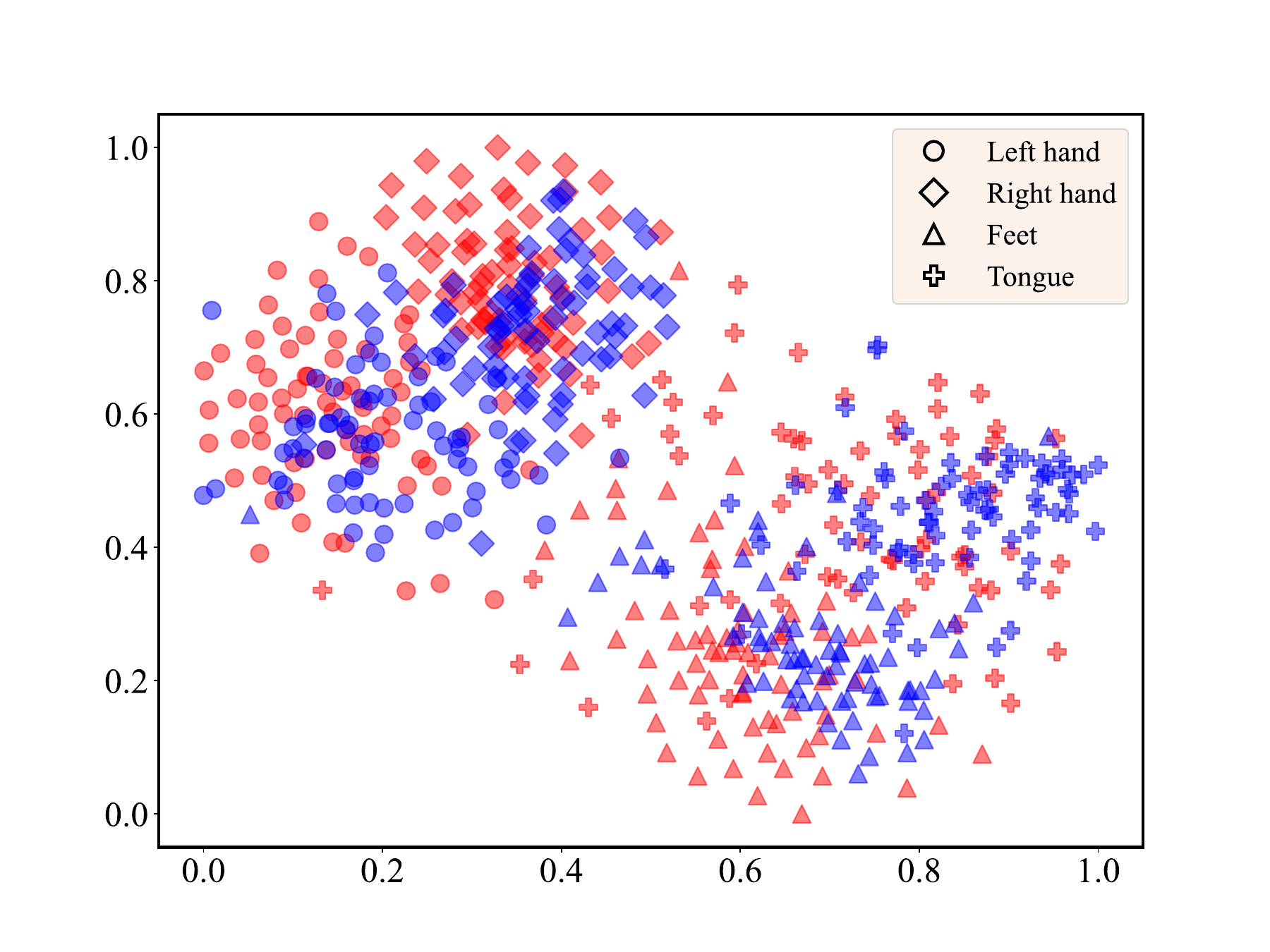}}
    \subfigure[Subject B05 before EEG-DG]{\includegraphics[scale = 0.22,trim=50 40 50 50,clip]{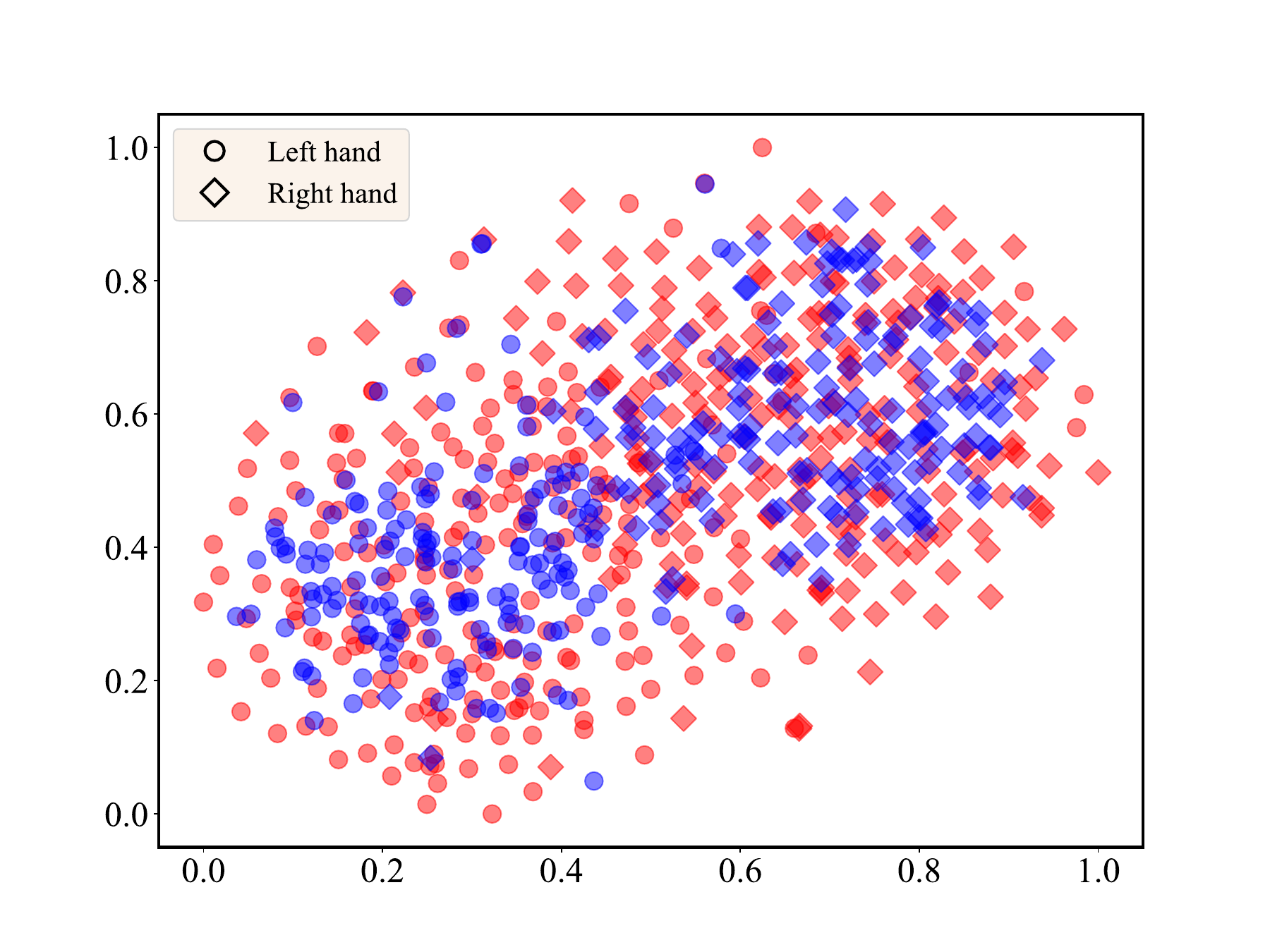}}
    \subfigure[Subject A03 after EEG-DG]{\includegraphics[scale = 0.22,trim=50 40 50 50,clip]{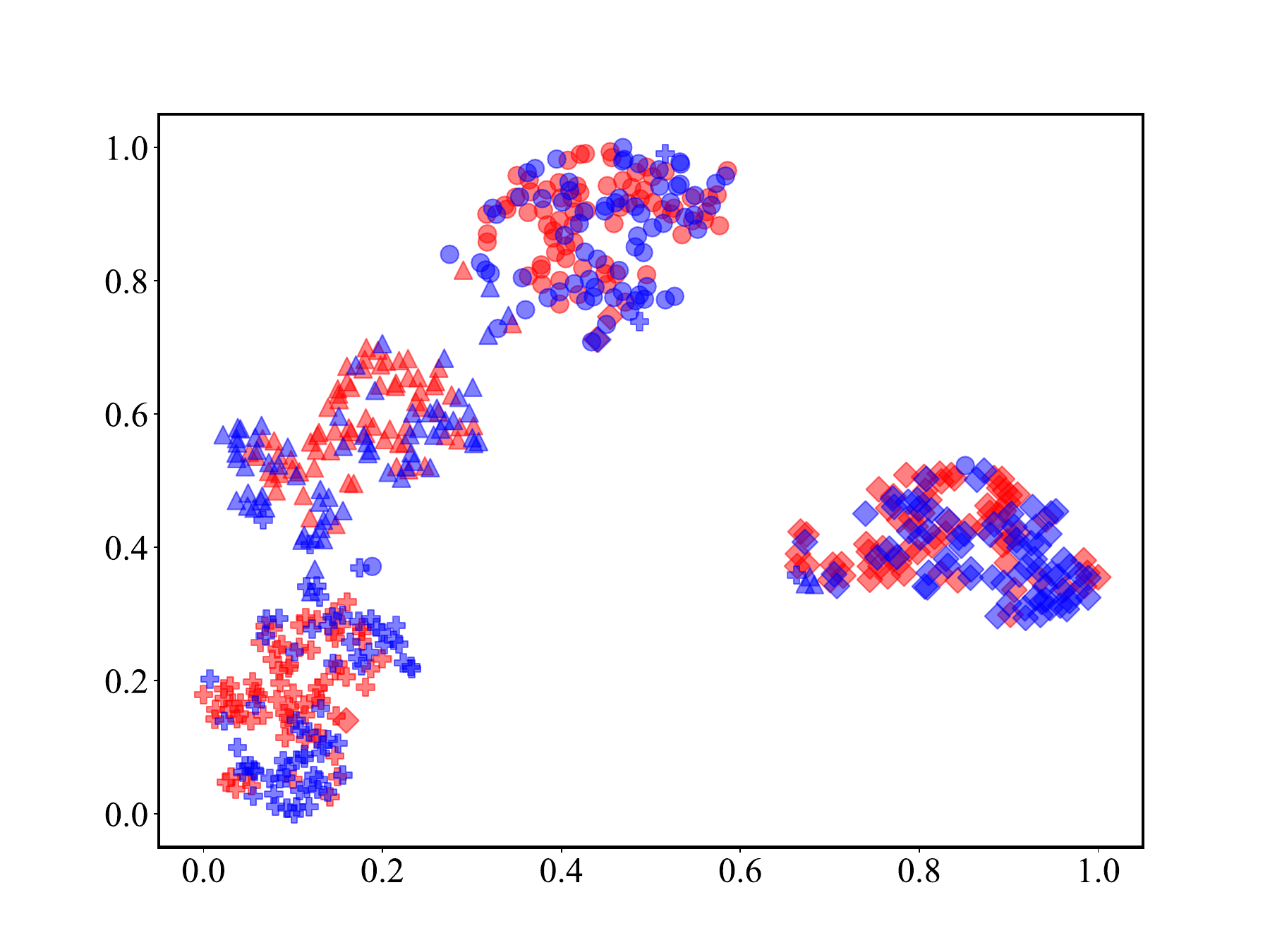}}
    \subfigure[Subject A07 after EEG-DG]{\includegraphics[scale = 0.22,trim=50 40 50 50,clip]{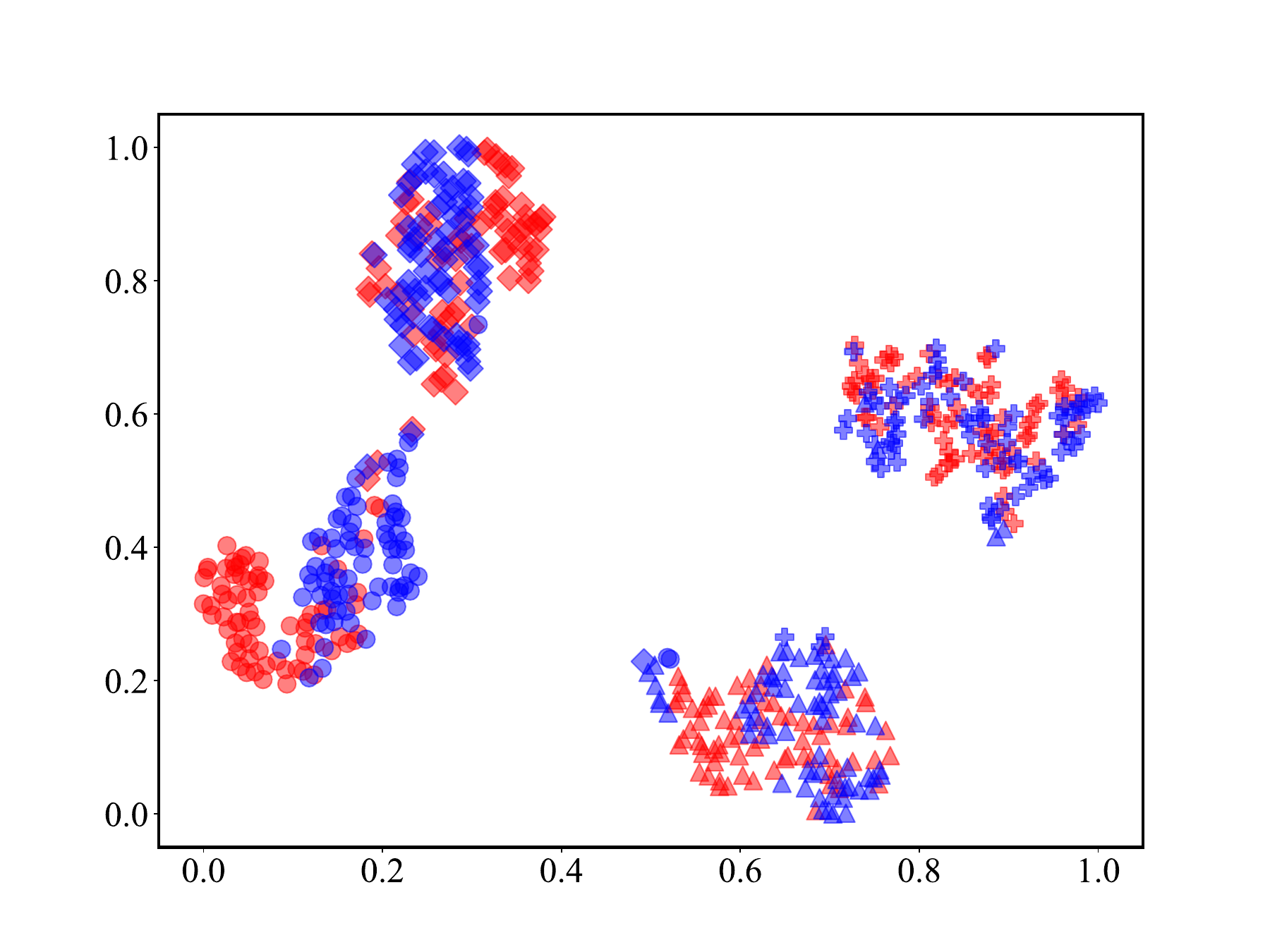}}
    \subfigure[Subject B05 after EEG-DG]{\includegraphics[scale = 0.22,trim=50 40 50 50,clip]{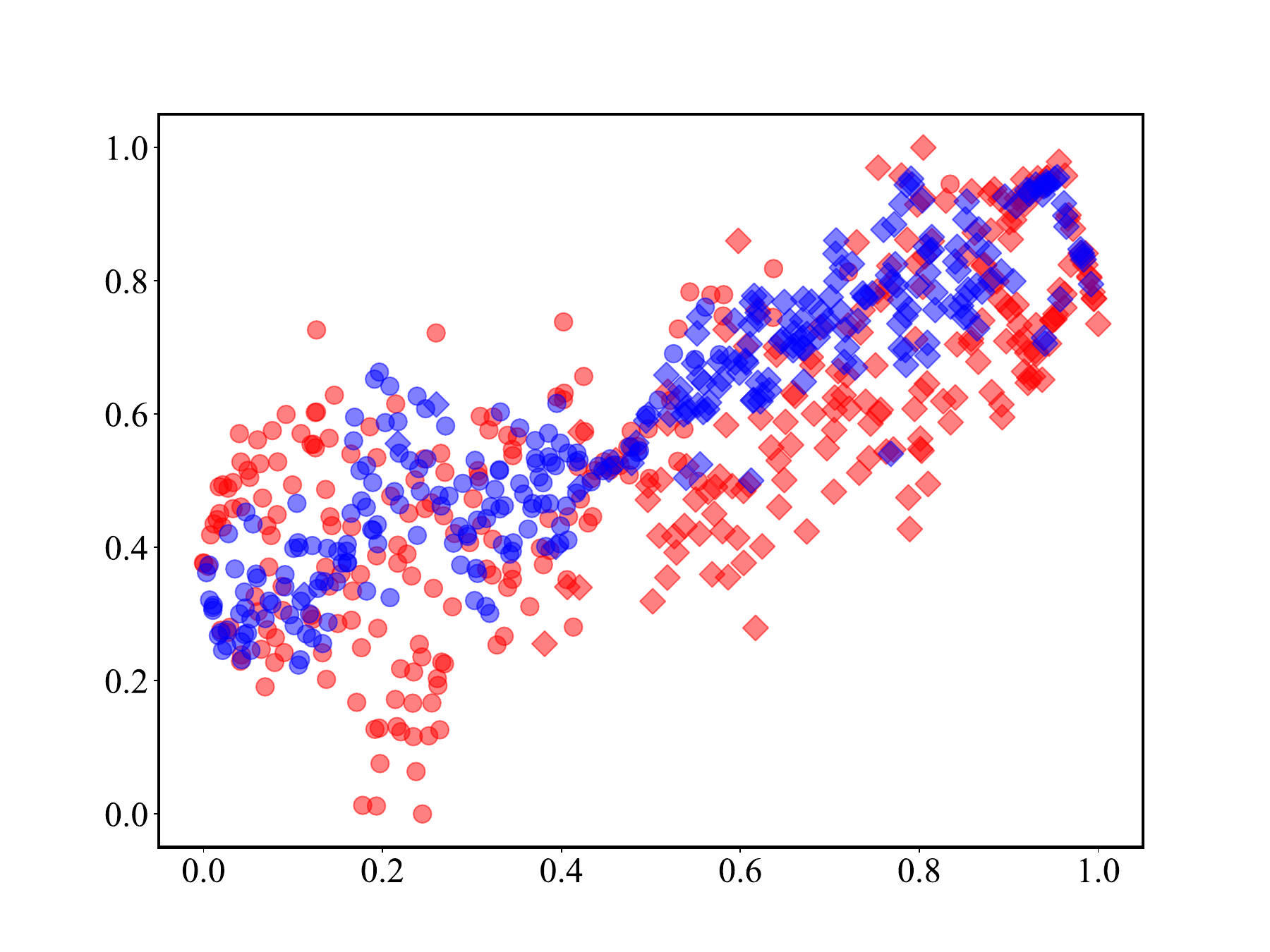}}
    \caption{Feature visualization with t-SNE. The top row displays the features extracted by the multi-scale feature extractor, while the bottom row shows the features obtained through the EEG-DG with domain-invariant feature representation. The figure employs different shapes to represent different motor imagery movements, with the red and blue colors indicating the source and target domains, respectively.}
    \label{fig_visual}
\vspace{-0.4cm}
\end{figure*}

\vspace{-0.1cm}
\subsection{Classification Performance Comparisons}
\vspace{-0.2cm}
The classification accuracy (in percentage \%), average classification accuracy (MEAN), standard deviation (STD), and kappa value for each subject on datasets IV-$2$a and IV-$2$b are tabulated in Tables \ref{table_2a} and \ref{table_2b}, respectively. The highest accuracy and kappa value are bold-faced. 
From Table \ref{table_2a}, it can be seen that deep learning methods EEGNet, ConvNet,
and C2CM show an overall improvement of approximately 5\% in classification accuracy compared to FBCSP and CCSP, indicating that deep learning models are able to learn more discriminative features than traditional methods for EEG classification. By adding domain adaptation to deep learning models, the deep domain adaptation methods obtain the expected enhancement, ranging from 0.29\% to 8.81\% in the average classification accuracy. Furthermore, DJDAN achieves a higher classification accuracy of 1.43\% than DAWD, which strongly demonstrates that reducing both marginal and conditional distribution differences is superior to considering only marginal distributional differences. 

We highlight in Table \ref{table_2a} that our proposed EEG-DG outperforms all the compared methods by a significant margin in terms of the average classification accuracy and kappa value. Our proposed EEG-DG surpasses the recently published GAT method by 5.21\% in terms of classification accuracy and boosts the kappa value by 0.1415.
In addition, compared to these deep domain adaptation methods, our proposed EEG-DG framework manages to learn domain-invariant features with the joint distribution alignment of multi-source domains in spite of the lack of target data. Note that our method still exceeds the current winner DJDAN with an improvement of 2.76\% and 0.1234 in classification accuracy and kappa value, respectively. On different subjects, we find that our proposed EEG-DG further improves its performance than other methods. Although the effectiveness is somewhat limited when dealing with subjects A02, A05, and A06 who are lowly separable, the proposed EEG-DG is the best performer on subjects A01, A03, A04, A07, A08, and A09.

From Table \ref{table_2b}, it is noteworthy that the classification performance of our proposed EEG-DG also far surpasses the state-of-the-art methods on average, reaching a classification accuracy of 87.12\% and a kappa value of 0.7424. 
Our result also exhibits a relatively smaller standard deviation of 10.89. Furthermore, our proposed EEG-DG outcompetes the traditional and deep learning methods almost on all different subjects. Except on B02 and B07, it is moderately inferior to the EEGNet. Compared to the five domain adaptation approaches, although our proposed EEG-DG is not the best on different subjects, it often takes second and third places, which is already well beyond our expectations.
It is obvious that the utilization of target data makes domain adaptation methods more competitive. While our proposed EEG-DG has no privilege over the target domain, it learns feature representations from multiple source domains that are also enjoyed in the target domain. 

\vspace{-0.3cm}
\subsection{Feature Visualization}
To visually illustrate the effectiveness of the EEG-DG in achieving domain-invariant feature representation, Figure \ref{fig_visual} presents the t-SNE visualization \cite{van2008visualizing} of feature distributions from the source and target domains before and after applying EEG-DG, respectively. 
The subjects A$03$, A$07$, and B$04$ are randomly chosen for illustration.
For visual clarity, different motor imagery movements are depicted using different shapes, and the source and target domains are highlighted in different colors, respectively.
From Figures \ref{fig_visual}(a), (b), and (c), it can be observed that samples from the same class exhibit clustering behavior, while the boundaries between different classes are particularly fuzzy, presenting varying degrees of confusion.
Comparing the top and bottom rows of the figure, we notice that after utilizing EEG-DG, the statistical distributions of the same class but different domains become more concentrated, and their boundary is clearer. According to the area of the clusters and the overlap of different domains in Figure \ref{fig_visual}(d), (e), and (f), we conclude that samples of the same class from different source domains as well as the target domain are all well aligned. This visually demonstrates that the proposed EEG-DG framework, which utilizes both the marginal distribution alignment and conditional distribution alignment, can effectively achieve domain-invariant feature representation.  
\vspace{-0.3cm}
\subsection{Ablation Study}
\begin{figure}[t]
\vspace{0.2cm}
    \centering
    \includegraphics[scale = 0.40,trim=15 600 40 625,clip]{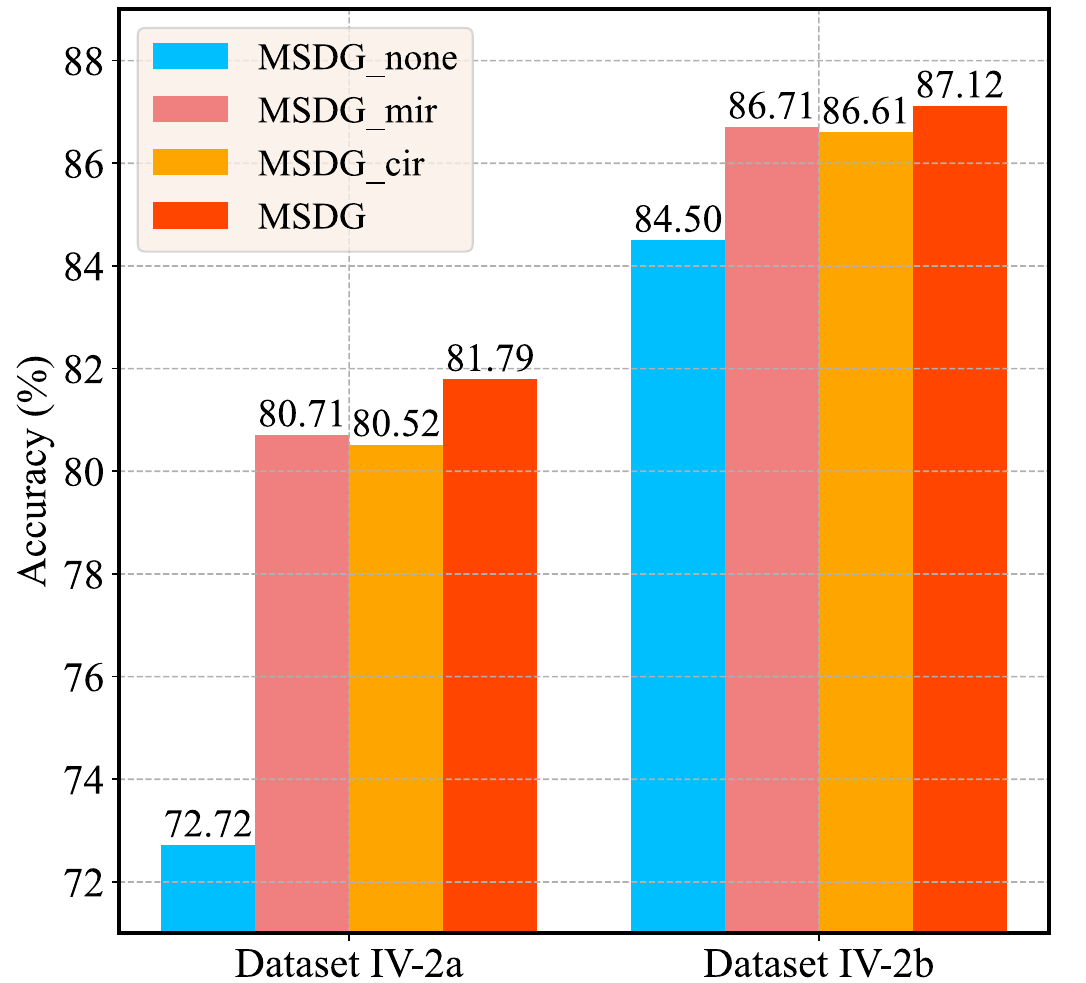}
    \caption{Classification accuracy of ablation experiments on BCI competition IV datasets IV-2a and IV-2b.}
    \label{fig_ablation}
\vspace{-0.4cm}
\end{figure}

\vspace{-0.1cm}
In our proposed EEG-DG, we aim to simultaneously reduce differences in marginal and conditional distributions across source domains to achieve domain-invariant feature representation. This is achieved through the optimization of the losses $\mathcal{L}_{mir}$ and $\mathcal{L}_{cir}$, each playing their respective roles. To demonstrate the effectiveness of optimizing each loss in EEG-DG, we perform the following ablation experiments. The results are presented in Figure \ref{fig_ablation}, where EEG-DG$_{mir}$ and EEG-DG$_{cir}$ represent the EEG-DG method with only margin-invariant representation ($\beta_1\neq0$ and $\beta_2=0$) and the EEG-DG method with only condition-invariant representation ($\beta_1=0$ and $\beta_2\neq0$), respectively. EEG-DG$_{none}$ refers to the EEG-DG method with only the feature extractor for EEG classification ($\beta_1=0$ and $\beta_2=0$), while EEG-DG represents the EEG-DG method with both margin-invariant representation and condition-invariant representation ($\beta_1\neq0$ and $\beta_2\neq0$).

From Figure \ref{fig_ablation}, it can be seen that compared to EEG-DG$_{mir}$ and EEG-DG$_{cir}$, the performance of EEG-DG$_{none}$ on dataset IV-2a is significantly inferior, which shows a decrease of 7.99\% and 7.80\% in terms of the average classification accuracy, respectively. When compared with EEGNet, EEG-DG$_{none}$ exhibits superior performance and surpasses EEGNet by 2.5\% due to the enhancements made on the multi-scale kernel. It confirms that the utilization of multi-scale convolutional kernels is beneficial for classifying EEG data with variations. Similar results are also reflected in dataset IV-2b. After incorporating the marginal distribution alignment and conditional distribution alignment, the overall classification performance is improved to varying degrees, while the improvement is more noticeable with the marginal distribution alignment.
It indicates that solely aligning the conditional distribution is less effective compared to solely aligning the marginal distribution, at the same time, simultaneously aligning both the marginal and conditional distributions is more advantageous for domain-invariant feature representation. 

\vspace{-0.2cm}
\section{Conclusion}
In this paper, we have proposed a multi-source domain generalization framework called EEG-DG to address the practical and challenging scenario where the target domain EEG data can not be accessed during training. This approach utilizes multiple source domains with different statistical distributions to construct generalizable models to make predictions on unseen target EEG data.
To reduce the distribution differences, our proposed EEG-DG achieves domain-invariant feature representation by simultaneously aligning both the marginal and conditional distributions across different domains.
Extensive experiments on a simulative dataset and BCI Competition IV-2a and IV-2b datasets have demonstrated the superiority of the proposed EEG-DG in handling EEG signals with non-stationary and individual variations against other state-of-the-art methods. 
However, one limitation of our method is that we currently divide the training data into different source domains solely based on the subject number or collection time. In practical scenarios where datasets, such as dataset IV-2a, lack collection time information, our method does not provide specific guidance for data division.
Our future research will focus on extending our EEG-DG to online transfer learning, enabling “plug-and-play" motor imagery BCI applications.

\ifCLASSOPTIONcaptionsoff
  \newpage
\fi

\bibliographystyle{IEEEtran}
\bibliography{Bibliography}

\vfill

\end{document}